\documentclass[12pt,preprint]{emulateapj}
\usepackage{aas_macros}
\usepackage{graphicx}
\usepackage{epstopdf}
\usepackage{verbatim}

\begin{document}

\title{Chondrule destruction in nebular shocks}

\author{Emmanuel Jacquet \& Christopher Thompson}
\affil{Canadian Institute for Theoretical Astrophysics, University of Toronto, 60 St George Street, Toronto, ON, M5S 3H8, Canada.}
\email{ejacquet@mnhn.fr}


\begin{abstract}
Chondrules are millimeter-sized silicate spherules ubiquitous in primitive meteorites, but whose origin remains mysterious. 
One of the main proposed mechanisms for producing them is melting of solids in shock waves in the gaseous protoplanetary disk. 
However, evidence is mounting that chondrule-forming regions were enriched in solids well above solar abundances. Given the 
high velocities involved in shock models destructive collisions would be expected between differently 
sized grains after passage of the shock front as a result of differential drag. We investigate the probability 
and outcome of collisions of particles behind a 1D shock using analytic methods as well as a full integration of the coupled mass, 
momentum, energy and radiation equations.  Destruction 
of protochondrules seems unavoidable for solid/gas ratios $\epsilon \gtrsim 0.1$, and possibly even for solar abundances because 
of ``sandblasting'' by finer dust.  A flow with $\epsilon \gtrsim 10$ requires much smaller shock velocities 
($\sim 2$ vs 8 km s$^{-1}$) in order to achieve chondrule-melting temperatures, and radiation trapping allows slow cooling of the
shocked fragments. Initial destruction would still be extensive;  
although re-assembly of mm-sized particles would naturally occur by grain sticking afterward, the compositional heterogeneity of 
chondrules may be difficult to reproduce.   We finally note that solids passing through small-scale bow shocks around few-km-sized 
planetesimals might experience partial melting and yet escape fragmentation.
\end{abstract}


\section{Introduction}

  A primary feature of nearly all primitive meteorites, or chondrites, is the presence of abundant millimeter-sized silicate spherules known as chondrules. They typically occupy $\sim$20-90 \% of the volume \citep{BrearleyJones1998}, and some debris of such objects have even been found in samples returned from comet Wild 2 \citep[e.g.][]{Bridgesetal2012}. Beyond all doubt, the high-temperature mechanism responsible for the formation of chondrules was a pervasive process in the early solar system, and likely other protoplanetary disks as well. Yet, despite two centuries of research since their original discovery \citep{Howard1802}, no consensus is in sight as to the nature of this mechanism \citep{Boss1996,ConnollyDesch2004,Ciesla2005,Krotetal2009,Deschetal2012}. A crucial first-order piece in the puzzle of protoplanetary disk physics is obviously missing.

  A leading contender in this vexed debate is melting by shock waves \citep[e.g.][]{Wood1963chondrules,HoodHoranyi1991,Iidaetal2001,DeschConnolly2002,Cieslaetal2004bow,Deschetal2005,BossDurisen2005,BossDurisen2005sources,MorrisDesch2010,Morrisetal2012,HoodWeidenschilling2012,Boleyetal2013,Nagasawaetal2014}. The basic picture is that if a portion of the disk is overrun by a sufficiently strong shock (say, $\gtrsim$ 6 km s$^{-1}$), the solids embedded in the gas 
will experience a strong drag, heating them to the point of melting. They rapidly approach thermal equilibrium with the hot (post-shock) gas, which cools as the shock front recedes away from them, allowing their eventual solidification. A consensus has not, however, been reached on a compelling mechanism for exposing primitive material to such shocks.  Amongst the various possibilities (see e.g. \citet{BossDurisen2005sources}), the debate has lately largely boiled down to gravitational instabilities \citep{BossDurisen2005,MorrisDesch2010} versus bow shocks produced by planetesimals \citep{Cieslaetal2004bow,HoodWeidenschilling2012} or planetary embryos \citep{Morrisetal2012,Boleyetal2013} on eccentric orbits. 

  The shock scenario can arguably boast a high level of theoretical development, with quantitative predictions on the thermal histories of chondrules comparing favorably with constraints from observations and furnace experiments, e.g. as to the 1-1000 K/h cooling rates inferred for them or the short time spent near the liquidus temperature \citep{Hewinsetal2005}. It must be noted though that recent bow shock simulations with radiative transfer have difficulties in reproducing the protracted cooling of porphyritic chondrules \citep{Boleyetal2013}. As for large-scale shocks such as those expected from gravitational instabilities, a ``pre-heating'' lasting hours until the shock is normally predicted, at variance with the lack of isotopic fractionation of S (a moderately volatile element) in putative primary troilite \citep{MorrisDesch2010,TachibanaHuss2005}; also \citet{StammlerDullemond2014} questioned whether sideway energy loss would be efficient enough to account for chondrule cooling rates if the shock scale is comparable to the disk thickness. 

  One important challenge to shock models was pointed out by \citet{NakamotoMiura2004} and \citet{Uesugietal2005}: ``protochondrules'' (chondrule precursors) of different sizes should decelerate (relative to the gas) at different rates in the post-shock region.  They would develop relative velocities commensurate with the full shock velocity (a few km s$^{-1}$), which would lead to their destruction upon collision
. \citet{Ciesla2006} examined the problem in some detail and argued that the \textit{molten} protochondrules would be able to withstand collision velocities up to a few hundred meters per second and the time span by which relative velocities would have decreased below that threshold would allow few collisions under canonical solid/gas ratios ($\sim 10^{-2}$; \citet{Lodders2003}).  

  Nonetheless, evidence is mounting that the chondrule-forming regions were considerably enriched in condensible elements relative to solar abundances. First, the FeO content of olivine in type I and especially type II chondrules, which is unlikely to have been inherited from nebular condensates \citep{Grossmanetal2012}, appears to record oxygen fugacities calling for solid/gas ratio of 10-1000 $\times$ solar \citep[e.g.][]{Schraderetal2013,FedkinGrossman2013}. Second, the significant proportion of compound chondrules---that is, pairs (or multiplets) of chondrules stuck together--- as well as the thickness of igneous rims, if interpreted to be accreted during chondrule formation \citep{Jacquetetal2013}, also point to solid densities typically a few orders of magnitude above those of ``standard'' solar nebula model \citep{Hayashi1981,Desch2007}, although this somewhat depends on the assumed collision velocities and cooling timescales \citep{GoodingKeil1981,Wassonetal1995,Cieslaetal2004,AkakiNakamura2005}. Third, the retention of Na in chondrules despite their high-temperature history \citep{Alexanderetal2008, Hewinsetal2012} indicates high partial pressures of Na, which, \textit{if} ascribed to partial volatilization from the chondrules themselves, would require solid concentrations more than five orders of magnitude above Minimum Mass Solar Nebula expectations \citep{Hayashi1981}. Fourth, in the specific framework of the large-scale shock models, an overabundance of fine dust may be needed to shorten the aforementioned pre-heating \citep{MorrisDesch2014}. In fact, some enhancement above solar may be theoretically expected anyway as a result of settling to the midplane or turbulent concentration \citep{Morrisetal2012}.

\begin{figure}
\plotone{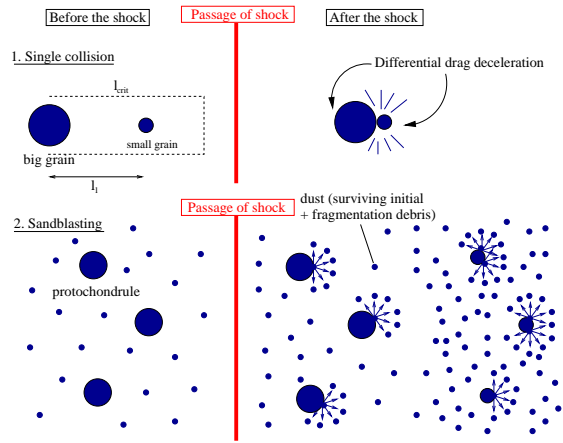}
\caption{Sketch of collisional destruction of chondrules in shocks. Time flows from left to right, with initial situation in the pre-shock region, passage of the shock front, and collisional evolution in the post-shock region. Solid particles are represented by blue spheres. The condition for collision is that the initial separation distance $l_1$ is shorter than a critical distance $l_{\rm crit}$ calculated in the text (``Single collision'', top). Chondrules may also be destroyed by continuous erosion through the impact of dust, whether surviving from the pre-shock region or produced \textit{in situ} by the very process of erosion (``Sandblasting'', bottom).}
\vskip .2in
\label{sketch}
\end{figure}

  However, enhanced solid/gas ratios would increase the probability of high-velocity collisions and may jeopardize anew the survival of chondrules. It is thus important to assess what maximum enrichment of the solid abundance is allowed in the shock model if wholesale destruction of chondrules is to be avoided. To that end, we investigate the dynamics and collisional evolution of solids through a gas-dominated shock, using analytic methods. Our focus is on the deceleration stage, independently of the source or large-scale structure of the shock. We are able to express the probability of collision between chondrules as a function of the solid/gas ratio, independently of the gas density. We likewise quantify the collisions of chondrules with finer dust, which may collectively lead to their near-total erosion, a process we will refer to as ``sandblasting''. This is depicted in Figure \ref{sketch}. 

  The difficulty we find in maintaining macroscopic particles behind the shock motivates a consideration of substantially different shock scenarios.  The settling of particles to the disk mid-plane can in some circumstances create large ratios of solids to gas.  We consider how the shocked flow is modified in such a situation and briefly consider the prospects for the reassembly of mm-sized particles after solids and gas equilibrate, and for their longer term survival.  Chondrule survival may also be enhanced in small-scale shocks.

  The outline of the paper is as follows: In Section \ref{Single}, we describe the dynamical and thermal evolution of a single particle, before envisioning the probability, velocity, and outcome of collisions in Section \ref{Collisions}. These two sections consider a gas-dominated shock. We will then consider solid-dominated shocks (Section \ref{Dusty shock}) and small planetesimal bow shocks where deceleration is incomplete (Section \ref{Bow}). In Section \ref{Discussion}, we summarize our results and discuss their implications on the viability of the shock model. Table \ref{Table symbols} lists the symbols used in the text.

\begin{table}
\caption{List of variables used in the text}
\label{Table symbols}
\begin{tabular}{c c}
\hline \hline
Symbol & Meaning \\
\hline
$a$ & Particle radius\tablenotemark{a}\\
$c$ & Speed of light\\
$c_g$, $c_{g,ad}$ & Isothermal, adiabatic gas sound speed\\
$c_s$ & Sound speed of dust-loaded gas\\
$C_p$ & Heat capacity of solids\\
$e_m$ & Particle emissivity\\
$E_{\rm rad}$ & Radiative energy density\\
$k_B$ & Boltzmann constant\\
$l_1$ & Pre-shock separation between two particles\\
$l_{\rm crit}$ & Critical value of $l_1$ for post-shock collision\\
$l_{\rm cool}$ & Post-shock cooling distance\\ 
$m$ & Particle mass\tablenotemark{a}\\
$M$ & Mach number $u/c_{g,ad}$\\
$M_{\rm pl}$ & Planetesimal mass\\
$n$ & Particle number density\tablenotemark{a}\\
$P$ & Pressure\\
$P_{\rm coll}$ & Catastrophic collision probability\\
$R_{\rm pl}$ & Planetesimal radius\\
$t_{\rm coll}$ & Collision time\\
$t_{\rm cool}$ & Post-shock cooling time\\
$t_m$, $t_{\rm cpd}$ & Time of chondrule mergers, compound formation\\
$t_{\rm th}$ & Particle heating timescale\\
$T$ & Temperature\tablenotemark{a}\\
$T_{\rm rec}$ & Recovery temperature (from gas molecular heating)\\
$T_{\rm peak}$ & Peak particle temperature\\
$T_{\infty}$ & Asymptotic post-shock temperature\\
$u$ & Particle velocity (in shock rest frame)\tablenotemark{a}\\
$u_g$ & Gas velocity (in shock rest frame)\\
$u_1=u_{g,1}$ & Particle/gas velocity upstream of shock \\
$u_J$ & Gas velocity jump across shock $u_1-u_{g,2}$\\
$u_\infty$ & Asymptotic speed of equilibrated flow\\
$v$ & Particle velocity relative to gas\\
$v_f$ & Velocity constant entering ejection yield $Y$\\
$v_T$ & Thermal speed $\sqrt{8k_BT_g/(\pi\mu_g)}$\\
$V$ & $v_s/u_J$ upon collision\\
$x$ & Abscissa of test particle\\
$x_{\rm cpd}$ & Compound chondrule frequency\\
$X_1$ & Fraction of mass flux in solids, $\epsilon_1/(1+\epsilon_1)$\\
$Y$ & Ejection yield\\
$\alpha$ & Power law exponent of ejection yield\\
$\gamma$ & Adiabatic exponent of gas\\
$\Delta v$ & Collision velocity\\
$\epsilon$ & Solid/gas mass ratio\tablenotemark{a}\\
$\theta_{\rm drag}$ & Stopping time correction factor\\
$\theta_{\rm th}$ & Thermal exchange correction factor\\
$\eta$ & Molten chondrule viscosity\\
$\mu_g$ & Mean molecular mass\\
$\rho$ & Mass density\tablenotemark{a}\\
$\rho_0$ & Internal particle density\\
$\sigma$ & Surface tension of molten chondrules\\
$\Sigma$ & Surface density\tablenotemark{a}\\
$\sigma_{\rm SB}$ & Stefan-Boltzmann constant\\ 
$\tau$ & Stopping time\tablenotemark{a}\\
$\Omega$ & Keplerian angular velocity\\
\hline
\end{tabular}
\tablenotetext{1}{Additional subscripts: ``1''/``2'' = pre-/post-shock, 
``g'' = gas, ``b''/''s'' = big/small particles, ``p'' = protochondrules,
``d'' = dust.}
\vskip .2in
\end{table}

\section{Single-particle Dynamics}
\label{Single}

We consider a one-dimensional normal 
 shock in the gas, with subscripts ``1'' and ``2'' referring to the pre- and post-shock regions respectively.  In what follows, $\rho_g$ is the density of the gas, and $P$ and $T_g$ its pressure and temperature, respectively.   
We denote by $u_g$ the gas velocity in the frame where the shock is stationary, and by $u_1$ the combined velocity of gas and particles
upstream of the shock.  Then the gas velocity jump is $u_J=u_{g,1}-u_{g,2} = u_1 - u_{g,2}$.
In this section we investigate the post-shock fate of one spherical solid particle of internal density $\rho_0$ and radius $a$, originally co-moving with the gas, by studying its motion and then its thermal evolution. 


Fragmentation already has important effects in gas-dominated flows, where the solid/gas mass ratio $\epsilon$ is $\ll 1$.
We therefore set aside the backreaction of solids on the gas flow in the analytic approach of Sections \ref{Single} and \ref{Collisions}.
The the figures show numerical results based on the full two-fluid equations developed in Appendix \ref{Dusty shock app} and
Section \ref{Dusty shock}.

\subsection{Motion in the post-shock gas}
  
  Since the hydrodynamic jump (a few molecular mean free paths in width) is much narrower than either the particle collisional mean free path or the stopping length, 
the particle arrives in the post-shock region at the velocity of the pre-shock gas, that is, with a velocity relative to post-shock gas equal to the jump velocity $u_J$. The jump velocity is trans-sonic in the post-shock region: from the Hugoniot Rankine relations \citep[e.g.][]{Deschetal2005}, for $T_{g,2}$ evaluated immediately after the shock:
\begin{eqnarray}
\frac{u_J}{\sqrt{\gamma k_BT_{g,2}/\mu_g}} & = & \frac{2\left(1-1/M_1^2\right)}{\sqrt{\left(2\gamma-(\gamma-1)/M_1^2\right)\left(\gamma-1+2/M_1^2\right)}}\nonumber\\
&\approx & \sqrt{\frac{2}{\gamma(\gamma-1)}}
\end{eqnarray}
with $\gamma$ the ratio of specific heats (taken to be 7/5 in numerical applications) and $M_1\equiv u_1/\sqrt{\gamma k_BT_{g,1}/\mu_g}$ the Mach number of the shock, with $\mu_g=3.9\times 10^{-27}$kg the mean molecular weight, and where the last equation holds in the limit $M_1\gg 1$ (as we will also adopt in numerical applications).

   From now on, we place ourselves in the \textit{frame of the post-shock gas}
, which will turn out to be quite convenient to calculate the dynamics of the particles. In this frame, the drag force is:
\begin{equation}
m\frac{\mathrm{d}v}{\mathrm{d}t}=-m\frac{v}{\tau}
\label{Newton Epstein}
\end{equation}
with $v$ the velocity of the particle relative to the gas, $m$ the particle mass and $\tau$ the stopping time given by:
\begin{equation}
\tau=\theta_{\rm drag}\frac{\rho_0a}{\rho_gv_T}
\label{tstop}
\end{equation}
 with $v_T=\sqrt{8k_BT_g/(\pi \mu_g)}$ the thermal speed and $\theta_{\rm drag}$ a correction factor. For a perfect conductor\footnote{And further assuming a particle temperature $T_p=T_{g,2}$, which is not quite true (Section \ref{Thermal evolution}) but order-unity relative deviations only incur  $\sim 10\%$
deviations in $\theta_{\rm drag}$.},
$\theta_{\rm drag}$ is 0.55 immediately after the shock for our shock conditions, 
and increases to $\theta_{\rm drag}=\left(1+\pi/8\right)^{-1}=0.72$ in the subsonic regime \citep{Gombosietal1986}. It will thus be a fair approximation to take it constant, which we choose to be 0.65, its velocity-averaged value in the \citet{Gombosietal1986} framework\footnote{This actually ensures that the stopping length $\int_{0}^{u_J}\tau\mathrm{d}v$ is accurate (in this specific model), same will hold then for the collision probability calculated in Section \ref{Collision probability}.}. For $\rho_{g,1}=10^{-6}$ kg m$^{-3}$, as typically considered by \citet{DeschConnolly2002,Morrisetal2012,Boleyetal2013} and $T_{g,2}=2000$ K, $\tau_2=0.4\:\mathrm{min}$.

Thus, taking $x=0$ to correspond to the position of the shock front at $t=0$ when the particles crosses it, the velocity of the particle is:
\begin{equation}\label{eq:vt}
v=u_Je^{-t/\tau}
\end{equation}
and the abscissa (increasing in the direction of the flow):
\begin{equation}\label{eq:xt}
x=u_J\tau\left(1-e^{-t/\tau}\right)=\tau\left(u_J-v\right)
\end{equation}
hence a stopping length of $u_J\tau$.

\subsection{Thermal evolution}
\label{Thermal evolution}

A particle entering the post-shock region is heated by photons and gas molecules. The energy evolution of the particle temperature $T_p$ is governed by \citep{Gombosietal1986,Deschetal2005}:
\begin{eqnarray}
\label{eq:Tp}
\frac{4}{3}\pi a^3\rho_0C_p\frac{dT_p}{dt} &=&
4\pi a^2\Bigg[\frac{\gamma+1}{\gamma-1}\frac{\rho_gk_Bv_T}{8\mu_g} \theta_{\rm th}\left(T_{\rm rec}-T_p\right)\nonumber\\ 
&&\quad\quad\quad + e_m\left(\frac{c}{4}E_{\rm rad}-\sigma_{\rm SB}T_p^4\right)\Bigg].
\end{eqnarray}
Here $e_m$ is the particle emissivity, $E_{\rm rad}$ the radiative energy density
, $C_p = 1.3\;{\rm kJ \;kg^{-1}\;K^{-1}}$ the specific heat capacity \citep{MorrisDesch2010}, $T_{\rm rec}$ the ``recovery temperature'' and $\theta_{\rm th}$ a correction factor normalized to the subsonic regime. For strong shocks with $\gamma=7/5$, we have $T_{\rm rec}=1.9T_{g,2}$ and $\theta_{\rm th}=1.7$ (see \citet{Gombosietal1986}).  

For penetration distances shorter than the photon mean free path, the radiation field is not significantly different from that in the pre-shock region and heating will be dominated by molecular collisions. Hence the heating timescale may be expressed as:
\begin{equation}
t_{\rm th}=\frac{8}{3}\frac{\gamma-1}{\gamma+1}\frac{\mu_g\rho_0aC_p}{\theta_{\rm th}\rho_gv_Tk_B} 
\end{equation}
Its ratio to the stopping time is then
\begin{equation}
\frac{t_{\rm th}}{\tau}=\frac{8}{3\theta_{\rm th}\theta_{\rm drag}}\frac{\gamma-1}{\gamma+1}\frac{\mu_gC_p}{k_B},
\end{equation}
that is 0.15 for our shock parameters. 
The particles thus reach peak temperature after a few tenths of their stopping time from passage through the shock
(as is demonstrated in the numerical calculation shown in Figure \ref{fig:temp} and discussed further in Section \ref{Dusty shock}).
If this temperature is in the chondrule-forming range, then the particles are at least partly molten from that point onward. 

The peak temperature may be obtained by balancing gas-particle heating and emission of radiation, 
\begin{eqnarray}
\label{Tpeak}
T_p &= &\left(\frac{\theta_{\rm th}T_{\rm rec}T_{g,2}^{1/2}\rho_{g,2}}{\sqrt{8\pi}\sigma_{\rm SB}e_m}\frac{\gamma+1}{\gamma-1}\right)^{1/4}\left(\frac{k_B}{\mu_g}\right)^{3/8}\\
&\approx & \left(\frac{\theta_{\rm th}\left(T_{\rm rec}/T_{g,2}\right)\rho_{g,1}u_1^3}{\sqrt{\pi\left(\gamma-1\right)}\left(\gamma+1\right)\sigma_{\rm SB}e_m}\right)^{1/4}\nonumber\\
&=& 1400~{\rm K}\left(\frac{u_1}{8\:\rm km~s^{-1}}\right)^{3/4}\left(\frac{\rho_{g,1}}{10^{-6}\:\rm kg~m^{-3}}\right)^{1/4}\nonumber\\
&&\times\left(\theta_{\rm th}\frac{T_{\rm rec}}{T_{g,2}}\right)^{1/4}\left(\frac{0.8}{e_m}\right)^{1/4},\nonumber
\end{eqnarray}
where the last two equalities refer to the strong shock limit.

\subsection{Evaporation}

With such high temperatures, evaporation may threaten the survival of the smaller particles \citep{DeschConnolly2002,MiuraNakamoto2005,MorrisDesch2010}. In order to assess their possible collisional role (in Section \ref{Erosion}), we need to investigate how much mass they lose to evaporation during chondrule deceleration.

  We first note that while the actual \textit{peak} temperature of the grain would correspond to a $\left(\theta_{\rm th}T_{\rm rec}/T_{g,2}\right)^{1/4}\approx 1.3$ enhancement relative to the $\theta_{\rm th}T_{\rm rec}/T_{g,2}=1$ normalization in Equation (\ref{Tpeak}), that temperature would hold for a timescale  $\sim\tau\propto a$.  The thickness evaporated during that particular time would thus be proportional to the radius (for grain sizes larger than the wavelength of emission $\sim 1\:\mu$m). So if the shock event is modeled as to allow chondrule survival during deceleration, dust should withstand that phase as well, and thus its survival on the \textit{chondrule} stopping time should be assessed with the temperature estimates corresponding to $\theta_{\rm th}T_{\rm rec}/T_{g,2}=1$.  We conclude that if the peak temperatures were in the range 
1400-1850$^\circ$C, as inferred for chondrules \citep{Hewinsetal2005}, the temperatures to consider should be $\lesssim 1700\:\rm K$.

  What are then the constraints on the evaporation rates, say for forsterite (the magnesian endmember of olivine, Mg$_2$SiO$_4$), a major chondrule-forming mineral?   The experimentally calibrated model of \citet{Tsuchiyamaetal1999} implies that the evaporation rate remains below $10^{-3}\:\mathrm{mol\:m^{-2}s^{-1}}$
at temperatures $< 1800\:\mathrm{K}$, given a total pressure $<10^{-3}\:\mathrm{bar}$ and solar abundances. The corresponding linear rates are $<0.04\:\mu$m s$^{-1}$, increasing by an order of magnitude if the temperature bound is taken to be 2000 K instead.  Dust larger than a micron should therefore survive deceleration
on the timescale $\tau_2=0.4\:\mathrm{min}$ corresponding to $\rho_{g,1}=10^{-6}\:\mathrm{kg~m^{-3}}$.
The same conclusion holds at smaller densities as well since the increase in stopping time is partly compensated by the decrease of the evaporation rate $\propto P_{\rm H_2}^{1/2}$ \citep{Tsuchiyamaetal1999}.

 Moreover, in a medium enriched in condensibles (as may be required by chondrule data, see Introduction), the evaporation enters the H$_2$O/H$_2$ buffer-dominated regime.  Here the evaporation rate is
suppressed by a factor $P_{\rm H_2}/P_{\rm H_2O}$ (see equation 9 of \citet{Tsuchiyamaetal1999}).  In fact forsterite becomes \textit{stable} over a wider temperature range, with e.g. complete evaporation at $\sim 2000$ K instead of 1400 K at $10^{-3}$ bar for dust/gas ratio of 1000 x solar \citep{Tsuchiyamaetal1999}).  So while submicron-sized dust, being a poorer radiator of heat than its bigger counterparts, would likely evaporate \citep{MorrisDesch2010}, supermicron-sized dust 
 may actually survive \textit{during the deceleration phase}.  This would not prevent complete evaporation before significant cooling has taken place further downstream:  for example, \citet{MiuraNakamoto2005} found that grains smaller than $\sim 10\:\mu$m would eventually completely evaporate. 

\section{Collisions}
\label{Collisions}

Having studied the fate of an isolated solid particle, we now consider the possibility of collisions between populations of grains of different sizes.  A given population (labelled ``$x$'') has a mean density $\rho_x$ and a solid/gas ratio $\epsilon_x=\rho_x/\rho_g$ (which will vary across the shock), and we designate by $\epsilon=\sum_{x}\epsilon_x$ the overall solid/gas ratio.
 We will first study the collision probability and velocity for a general two-size population. Following a discussion of the outcome of the collisions, we evaluate the extent of catastrophic fragmentations before turning to the possibility of progressive ``sandblasting'' of protochondrules by dust.  We finally assess the possibility of recoagulation further downstream. 
  
\subsection{Probability and speed of collision between \\ two grains of different sizes}
\label{Collision probability}


In this subsection, we calculate the probability of collision of a given big grain (henceforth with subscript ``$b$'') with a member of a population of small grains (subscript ``$s$'') originally co-moving with it. To that end, we seek the maximum initial separation $l_1$ of a small grain downstream of a big grain that will allow it to be overrun by the big grain in the post-shock region. Knowing the collisional cross section and the density of the small particles then allows us to obtain the collision probability for the big grain in closed form. The only restriction made here is that the shock is gas-dominated, i.e. $\epsilon_1 \ll 1$.

  We adopt the convention that $t=0$ and $x=0$ corresponds to the entry of the \textit{smaller} particle in the post-shock region (remember we work here in the frame of the post-shock gas). Taking into account the fact that the bigger grain enters the post-shock region at time $t=l_1/u_1$ and at abscissa $x=-u_{g,2}l_1/u_1$ (as the shock front is receding backward in the frame of the post-shock gas), the time of collision $t_{\rm coll}$ --- if it does take place --- may be calculated by setting the abscissas equal:
\begin{eqnarray}
u_J\tau_s\left(1-e^{-t_{\rm coll}/\tau_s}\right) &=& -\frac{u_{g,2}}{u_1}l_1 \nonumber\\
                 &+& u_J\tau_b\left(1-e^{-(t_{\rm coll}-l_1/u_1)/\tau_b}\right).\nonumber\\
\end{eqnarray}
This may be rewritten in terms of the variable $V \equiv v_s/u_J$ as
\begin{equation}
\label{V}
\frac{\tau_s}{\tau_b}-1+\frac{l_1u_{g,2}}{u_1u_J\tau_b}-\frac{\tau_s}{\tau_b}V+e^{\frac{l_1}{u_1\tau_b}}V^{\tau_s/\tau_b}=0.
\end{equation}
This equation has a (unique) solution $V\in ]0;1[$  -- that is, collision will actually occur -- if and only if:
\begin{equation}
\label{coll condition}
l_1<l_{\rm crit}\equiv \left(\tau_b-\tau_s\right)\frac{u_1u_J}{u_{g,2}}.
\end{equation}
This simply amounts to saying that the separation of the entry points in the post-shock region (in its rest frame) $u_{g,2}l_1/u_1$ must be smaller than the difference of stopping lengths $u_J\tau_b-u_J\tau_s$ between the big and the small particle. Introducing the collisional mean free path in the pre-shock region $l_{\rm coll,1}=(n_{s,1}\pi \left(a_b+a_s\right)^2)^{-1}$ with $n_s$ the number density of small grains, and given that the number of grains in a given volume obeys a Poisson distribution, we obtain a collision probability 
\begin{equation}
\label{Pcoll}
P_{\rm coll}=1-e^{-l_{\rm crit}/l_{\rm coll,1}}=1-e^{-\epsilon_{s,1}/\epsilon_{\rm s,crit}}
\end{equation}
with
\begin{equation}
\label{epsilon_crit}
\epsilon_{\rm s,crit}=\frac{4v_{T,2}}{3\theta_{\rm drag}u_J}\frac{\left(a_s/a_b\right)^3}{\left(\rho_{0,b}/\rho_{0,s}-a_s/a_b\right)\left(1+a_s/a_b\right)^2}.
\end{equation}
The first factor works out to 1.5 given our shock parameters.

If the condition (\ref{coll condition}) is met, what are the collision speeds? 
The collision velocity $\Delta v=v_b-v_s$ is obtained from Equations (\ref{eq:vt}) and (\ref{eq:xt}),
\begin{equation}
\frac{\Delta v}{u_J}=e^{\frac{l_1}{u_1\tau_b}}V^{\tau_s/\tau_b}-V=\left(1-\frac{\tau_s}{\tau_b}\right)\left(1-V-\frac{l_1}{l_{\rm crit}}\right),
\end{equation}
where we have used Equation (\ref{V}) in the second equality. Expressing $V$ as a function of $\Delta v$ and inserting in Equation (\ref{V}), 
we obtain a direct relationship between $l_1$ and $\Delta v$:
\begin{eqnarray}
1-\frac{\Delta v}{u_J\left(\tau_b/\tau_s-1\right)}-\frac{l_1}{l_{\rm crit}} = 
\quad\quad\quad\quad\quad\quad\quad\quad\quad\nonumber\\
 e^{\frac{l_1}{u_1\tau_b}}\left(1-\frac{\Delta v}{u_J\left(1-\tau_s/\tau_b\right)}-\frac{l_1}{l_{\rm crit}}\right)^{\tau_s/\tau_b}.
\end{eqnarray}

We can work out useful asymptotic solutions in two complementary regimes:

\vskip .1in
\noindent 1. For $l_1\ll u_J\tau_s\left(1-\tau_s/\tau_b\right)$, one finds $V=1-\sqrt{2l_1/(u_J\tau_s(1-\tau_s/\tau_b))}+O(l_1/(u_J\tau_s))$
(that is, the smaller particle has not completely decelerated).  The collision velocity is given by
\begin{eqnarray}
\frac{\Delta v}{u_J} &\approx & \sqrt{\frac{2l_1(1-\tau_s/\tau_b)}{u_J\tau_s}}\\
&=& \sqrt{\frac{8}{3\theta_{\rm drag}}\frac{v_{T,2}}{u_J}\frac{u_1}{u_{g,2}}}\left(\frac{1-a_s/a_b}{\epsilon_{s,1}}\right)^{1/2}\left(\frac{l_1}{l_{\rm coll,1}}\right)^{1/2}\nonumber\\
&& \times\left(\frac{a_s}{a_b+a_s}\right).\nonumber 
\end{eqnarray}
The factor with the square root in the last equality evaluates to 4.2 for our fiducial shock parameters. 

\vskip .1in
\noindent 2. For $u_J\tau_s\left(1-\tau_s/\tau_b\right)\ll l_1 \leq l_{\rm crit}$, one finds
$V\approx e^{-l_1/(u_1\tau_s)}\left(\left(1-\tau_s/\tau_b\right)\left(1-l_1/l_{\rm crit}\right)\right)^{\tau_b/\tau_s}$,
which is negligibly small (that is, the small particle has essentially been stopped by the gas).  Then
\begin{equation}
\label{Delta v s stopped}
\frac{\Delta v}{u_J}\approx \left(1-\frac{a_s}{a_b}\right)\left(1-\frac{l_1}{l_{\rm crit}}\right). 
\end{equation}
\vskip .1in

\begin{figure}
\plotone{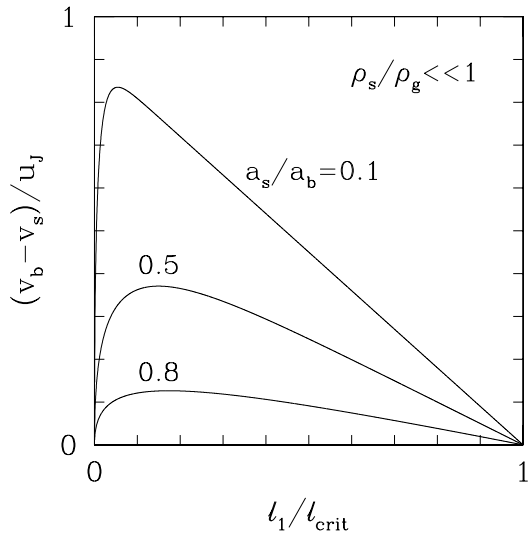}
\caption{Collision velocity (normalized to jump velocity) as a function of the initial separation $l_1$ of the big and the small particle, normalized to the maximum separation for collision $l_{\rm crit}$. We assume a density contrast of 6 (as in strong shocks with $\gamma=7/5$). The curves are drawn for three different values of the small/big particle size ratio $a_s/a_b$. Two regimes are apparent: for short $l_1$, the small particle is still decelerating just before collision while for long $l_1$, the small particle is essentially stopped in the post-shock gas and swept by the still decelerating bigger particle.}
\vskip .2in
\label{collvel l1}
\end{figure}

\begin{figure}
\plotone{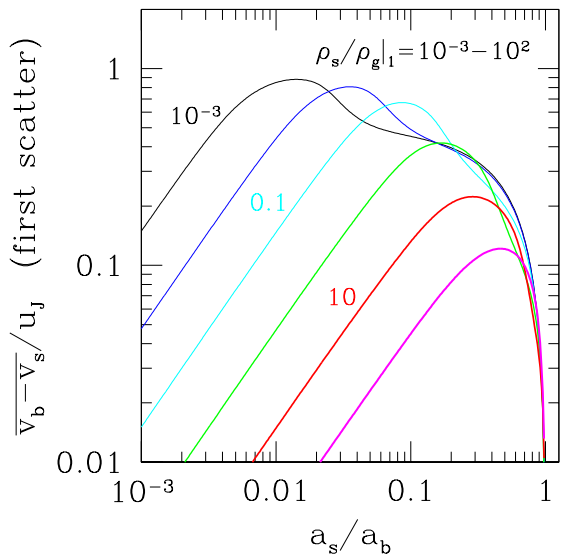}
\caption{Mean speed of first collision (normalized to jump velocity) as a function of size ratio $a_s/a_b$.  Strong shock
in $\gamma = 7/5$ gas, with solid/gas ratio $\epsilon_1 = 10^{-3}$-$10^2$.   For $\epsilon_1 \gtrsim 1$ the post-shock flow is
modified by the interaction of gas and solids.  We use the two-fluid equations of Section \ref{Dusty shock}, combined with the prescription for erosion given in Equations (\ref{Y}) and (\ref{dmdt}).  (Parameters chosen are $v_f = 0.04 u_1$ and relative mass fractions $(0.9,0.1)$ in big and small particles.)}
\vskip .2in
\label{collision speed}
\end{figure}

\begin{figure}
\plotone{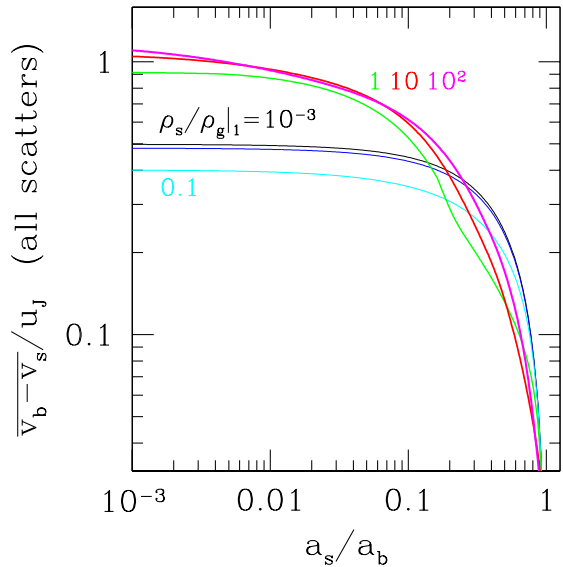}
\caption{Same as Figure \ref{collision speed}, but now averaged over all collisions.} 
\vskip .2in
\label{collision speed2}
\end{figure}

The general collision velocity as a function of $l_1$ is plotted in Figure \ref{collvel l1}. It reaches a maximum at
$l_1\sim u_J\tau_s\left(1-\tau_s/\tau_b\right)$ and decreases both for short $l_1$ (the two particles have had less time to 
develop relative velocities downstream of the shock), and for long $l_1$ (both particles are then increasingly coupled to the gas). 
We also plot the mean first collision speed in Figure \ref{collision speed} and the collision speed averaged over all collisions in Figure 
\ref{collision speed2}, as a function of the size ratio for different solid/gas ratios. 
Clearly, $\Delta v$ is typically within one order of magnitude of the full jump velocity.



\subsection{Outcome of collisions}
\label{Outcome of collisions}

Collisions between similar-size silicate aggregates at speeds in excess of $\sim 1$ m s$^{-1}$ will lead to 
fragmentation \citep{Guettleretal2010}; the critical speed may rise to $\sim 10$ m s$^{-1}$ in the case of less 
porous targets and/or stickier material like water ice or organics \citep{Guettleretal2010} or to $\sim 100$ m s$^{-1}$ 
for entirely coherent bodies \citep{Vedderetal1974}. The latter is however unlikely to apply to chondrule precursors 
(except in the case of re-melting of preexisting chondrules) given the absence of such coherent bodies in chondrites -- 
apart from the chondrules themselves and rare and compositionally distinctive igneous CAIs and other clasts.
In the case of unequal size collisions (a small projectile hitting a bigger target), fragmentation of the target would not be necessarily catastrophic.
Using dimensional analysis \citep{Holsapple1993}, the ratio of the ejected mass to the projectile mass, which we will call the ``ejection yield'' may be cast in the form:
\begin{equation}
\label{Y}
Y(\Delta v)=\left(\frac{\Delta v}{v_f}\right)^\alpha.
\end{equation}
Here $v_f$ is a constant with the dimensions of velocity, and the index $\alpha$ is expected to lie between 1 and 2 (limits corresponding to ejected mass proportional to the incident momentum and energy, respectively: \citealt{Housenetal1983,Holsapple1993,Poelchauetal2013}).
Hypervelocity impact experiments using a variety of target materials (wet soil, soft rock, hard rock) yield $\alpha=1.65$ in the strength regime, with $v_f=0.06-0.24$ km s$^{-1}$;  whereas for sand $\alpha=1.23$ and $v_f=5\:\rm m$ s$^{-1}$ (\citealt{Holsapple1993}; see also the recent MEMIN experiments: 
\citealt{Poelchauetal2013}). In either case,  we would have $Y(\Delta v)\sim 10^2-10^3$ at collision speeds $\Delta v$ of order a few km s$^{-1}$. 

 For collisions occurring after melting was complete, new - albeit less well understood physics -- would come into play \citep{QianLaw1997,Planchetteetal2012}.
 Under the assumption that breakup is controlled by some critical value ($\sim 10-100$) of the Weber number $We\equiv 2\rho_0a\Delta v^2/\sigma$ with $\sigma$ the surface tension, \citet{Kring1991} derived critical fragmentation velocities commensurate with those for solids ($\sim 1$ m s$^{-1}$). \citet{Ciesla2006} then argued that viscosity, which may vary by orders of magnitude during cooling, may make droplets more robust against disruption. It must be commented though that inasmuch as ferromagnesian chondrules have \textit{basaltic} compositions, they should have relatively low viscosities around liquidus temperatures: bulk chondrule viscosity estimates by \citet{GoodingKeil1981viscosity} only exceed 1 Pa.s around 1700 K. 
The viscosity would be increased in the presence of crystals, both by physical \citep{Roscoe1952} and melt-compositional \citep{GoodingKeil1981viscosity} effects, but this would be initially limited ($<$ 1-2 orders of magnitude)
. Also, experiments of drop impact on liquid films suggest that the critical splashing velocity has a weak dependence on viscosity
 \citep{Walzel1980, Rein1993,Yarin2006}. Using the expression given by \citet{Walzel1980}, the splashing condition is:
\begin{equation}\label{eq:vtension}
\Delta v>23\,\mathrm{m~s^{-1}}\left(\frac{\mathrm{kg~m^{-2}}}{\rho_0a}\right)^{0.6}\left(\frac{\eta}{\rm Pa~s}\right)^{0.2}\left(\frac{\sigma}{0.4\:\rm N~m^{-1}}\right)^{0.4},
\end{equation} 
with $\eta$ the (dynamic) viscosity. So it would seem that collision velocities above a few $\times 10$ m s$^{-1}$ should lead to disruption---a somewhat lower threshold range than envisioned by \citet{Ciesla2006}.   

Little seems to be known about the velocity-scaling of the ejection yield at small ratios of projectile to target mass.  
Micron-sized grains impacting on a melt surface will produce splash at speeds above $\sim 0.5$ km s$^{-1}$ according to Equation (\ref{eq:vtension}).
In the gravity-dominated regime (irrelevant here), the behavior of water is fairly similar to that of solid substrates, though crater volumes are more than one order of magnitude larger at the same impact energy \citep{Holsapple1993}. 
We will continue to adopt the general form of Equation (\ref{Y}) for the ejection yield,
 but it should be borne in mind that experimental data are still lacking for droplet collisions, as previously emphasized by \citet{Ciesla2006}.

  A complication which we have ignored is the dependence of collisions on angle (or equivalently on impact parameter).  If we were to consider that the general ejection yield is given by the same expression as Equation (\ref{Y}) but with $\Delta v$ replaced by its normal component \citep[e.g.][]{HousenHolsapple2011}, the yield averaged over the whole cross section would be its maximum value divided by $1+\alpha/2$. Whatever the angular dependence may actually be, it would likely only yield a small correction compared to the global uncertainty in $Y$.

\subsection{Critical solid abundances for \\ extensive catastrophic collisions}
\label{Critical solid abundance}

  What solid/gas thresholds for extensive catastrophic collisions do we obtain? From Section \ref{Collision probability}, we have seen that collisions, if they take place, occur at velocities comparable to $u_J$. A particle of size $a_s\gtrsim a_b\left(\rho_{0,b}/\rho_{0,s}\right)^{1/3}/Y(u_J)^{1/3}$ would thus typically incur catastrophic disruption of the big particle upon collision. If we take $Y(u_J)\sim 10^3$ this translates into a critical $a_s/a_b\approx 0.1$, and from Equation (\ref{Pcoll}), such (catastrophic) collisions will be frequent for $\epsilon_{s,1}\gtrsim 10^{-3}$, a threshold lower than the solar total solid/gas ratio.
  
 Yet individual components of such size are a minority in chondrites, with observed chondrule size distributions being quite narrow (typically within a factor of 2; e.g. \citealt {Cuzzietal2001,NelsonRubin2002}).  The size distribution of solids may, of course, have been different in the chondrule-forming region: for instance, \citet{MorrisDesch2014} considered a population of $\sim$10 $\mu$m grains---eventually evaporated---to alleviate the problem of pre-heating in large-scale shocks---this however would, as we see, exacerbate that of collisional destruction. But a conservative approach may be to restrict attention to the presently observed size distribution. If we consider the size distribution of metal grains measured by \citet{Guignard2011} for the Forest Vale H4 chondrite, generalizing Equation (\ref{Pcoll}) to a polydisperse population (by replacing the argument of the exponential with an integral over the size distribution), we obtain $\epsilon_{\rm s,crit}=0.005$, which, considering that metal only makes up 5.37 wt\% of this meteorite, amounts to a critical \textit{total} solid/gas ratio of 0.1. A similar threshold can be obtained from considering collisions with chondrules (which, although representing the majority of the mass, offer smaller size contrasts); if we take $\rho_{0,s}=\rho_{0,b}$ and $a_s/a_b=1/2$, we obtain $\epsilon_{\rm s,crit}=0.16$. So catastrophic collisions, while rare for solar abundances, as seen by \citet{Ciesla2006}, should be prevalent for solid/gas ratios $\gtrsim 10^{-1}$. 

  Given that the solid/gas ratio is high enough to induce catastrophic disruptions of mm-sized particles, might the chondrules we observe derive from fragments of much bigger precursors?  There are several arguments against such a possibility.  First, it is difficult theoretically to achieve sizes larger than mm or cm by coagulation in the disk \citep[e.g.][]{Braueretal2008,Birnstieletal2010,Zsometal2010}, except perhaps via the destruction of a prior generation of km-size gravitationally bound objects,
and indeed inclusions larger than mm are rare in chondrites.  Second, fragmentation would also be expected to produce a range of sizes, which is at variance with the narrow size distribution exhibited by chondrules.  Some narrowing of the size distribution may be possible following chondrule formation \citep[e.g.][]{Cuzzietal2001}, but \cite{Jacquet2014size} has argued against such a process on compositional grounds.  
Catastrophic disruption during the shock, if prevalent, would thus be difficult to reconcile with meteoritic evidence (see also Section \ref{Recoagulation}).

\subsection{Sandblasting}
\label{Erosion}

  Although collisions between protochondrules are violent, most collisions experienced by a given protochondrule will be with small dust particles, which could \textit{cumulatively} have a stronger effect. The dust in question may originally be (molten) dust surviving from the pre-shock region (even if restricted to its most refractory components in case of pre-heating), the debris of protochondrule-protochondrule collisions (representing a proportion $\sim P_{\rm coll}$ of the solids; see Equation (\ref{Pcoll})); or the result of ram pressure stripping of large droplets \citep{SusaNakamoto2002,Kadonoetal2008}.    
  
  Small particles need contribute only a modest fraction  (\ref{epsilon_crit}) of the mass flux in order to maintain frequent collisions
with larger grains.   We will therefore model fine dust by a continuous fluid (identified with subscript ``d''), which is coupled tightly to the gas on the short timescale (\ref{tstop}).  Collisions of dust grains with protochondrules (considered as a population of identical particles, identified with subscript ``p'') thus take place at the full speed $v$. 

 Under the influence of erosion, the mass $m_p$ of the protochondrule evolves as:
\begin{equation}\label{dmdt}
\frac{\mathrm{d}m_p}{\mathrm{d}t}=-\pi a_p^2\rho_dv Y(v).
\end{equation}
Dividing by Equation (\ref{Newton Epstein}) (where we neglect momentum transferred by the impinging dust as well as recoil from erosion) 
yields:
\begin{equation}
\label{dm/dv}
\frac{\mathrm{dln}m_p}{\mathrm{d}v}=\frac{3\theta_{\rm drag}\epsilon_d}{4v_{T,2}}Y(v).
\end{equation}
 The dust-to-gas ratio $\epsilon_d$ also increases at the expense of the eroded protochondrules\footnote{We assume here that newly produced dust is homogeneously distributed in order to back-react on the protochondrules. This requires that  fragments ejected at a velocity $v_{\rm ej}\approx u_J/Y(u_J)^{1/2}$ (from energy conservation) can reach the closest neighbours at a distance $n_p^{-1/3}$ [where $n_p=3\rho_p/(4\pi\rho_0a_p^3)$ is the number density of the protochondrules], within their stopping time $\tau_f$. This yields the condition:
\begin{eqnarray}
\frac{n_{p,1}^{-1/3}}{v_{\rm ej}\tau_f}&=&
\frac{u_1v_{T,2}}{u_{g,2}u_J}\frac{Y(u_J)^{1/2}}{\theta_{\rm drag}}\left(\frac{4\pi}{3\epsilon_{p,1}}\right)^{1/3}\left(\frac{\rho_{g,1}}{\rho_0}\right)^{2/3}\left(\frac{a_p}{a_f}\right)\nonumber\\
&=& 10^{-3}\left(\frac{a_p}{a_f}\right)\left(\frac{Y(u_J)}{10^2}\right)^{1/2}\left(\frac{\rho_{g,1}}{10^{-6}\:\mathrm{kg~m^{-3}}}\right)^{2/3}\left(\frac{10^{-2}}{\epsilon_{p,1}}\right)^{1/3}\nonumber\\
 \ll 1.
\end{eqnarray}
}:
\begin{equation}
\epsilon_d=\epsilon_{d,1}+\epsilon_{p,1}\left(1-\frac{m_p}{m_{p,1}}\right).
\end{equation}

After integration, we obtain\footnote{If we had considered a varying $\theta_{\rm drag}$ we would have had to replace it in the following equation by its $v^\alpha$-weighted velocity average, which, for our shock parameters and $\alpha=2$ would be 0.60 instead of our adopted 0.65 in the \citet{Gombosietal1986} framework---a fairly negligible correction.}
\begin{eqnarray}
\label{sandblasting result}
\frac{m_p}{m_{p,1}}
&=&\Bigg(\frac{\epsilon_{p,1}}{\epsilon_1}+\frac{\epsilon_{d,1}}{\epsilon_1}\nonumber\\
&\times& \mathrm{exp}\Bigg[\frac{3\theta_{\rm drag}u_J}{4v_{T,2}}\frac{\epsilon_1Y(u_J)}{\alpha+1}\left(1-\left(\frac{v}{u_J}\right)^{\alpha+1}\right)\Bigg]\Bigg)^{-1}\nonumber\\
\end{eqnarray}
with $\epsilon_1=\epsilon_{d,1}+\epsilon_{p,1}$ (see also Figure \ref{size}). 

  Equation (\ref{sandblasting result}) is equivalent to:
\begin{equation}
\epsilon_1=Y(u_J)^{-1}\frac{\alpha+1}{1-(v/u_J)^{\alpha+1}}\frac{4v_{T,2}}{3\theta_{\rm drag}u_J}\mathrm{ln}\left(\frac{\epsilon_1}{\epsilon_{d,1}}\frac{m_{p,1}}{m_p}-\frac{\epsilon_{p,1}}{\epsilon_{d,1}}\right).
\end{equation}

So quite insensitively to the seed dust/protochondrule ratio and the threshold $m_p/m_{p,1}$ below which a chondrule is conventionally considered destroyed, for a critical erosion velocity $v\ll u_J$, a destruction condition may be extracted as essentially 
\begin{equation}
\label{sandblasting condition}
 \epsilon_1\gg Y(u_J)^{-1}.
\end{equation}
 Yields  $Y(u_J) \sim 10^2-10^3$, as deduced in Section \ref{Outcome of collisions}, allow at best 
a limited enrichment of solids above solar composition ($\epsilon_1 \sim 10^{-2}$).
This constraint is more stringent than the one obtained from protochondrule-protochondrule collisions 
(Section \ref{Critical solid abundance}).

  Given the uncertainties in the rheology of liquid droplets (Section \ref{Outcome of collisions}), one may resort to a conservative approach considering only the initial phase where protochondrules are solid.  
(We assume that a liquid state of the impinging droplets does not qualitatively change their effect on the solids, as in a ``point source'' picture of the impact, e.g. \citealt{Holsapple1993}.) 

Then sandblasting would be restricted to velocities $\gtrsim(1-t_{\rm th}/\tau)u_J$, and the threshold for $\epsilon_1$ would need to be divided by $\left(1-(1-t_{\rm th}/\tau)^{\alpha+1}\right) \approx \left(\alpha+1\right)t_{\rm th}/\tau$.  The minimum solid fraction would be raised by almost an order of magnitude,  to  $10^{-2}-10^{-1}$ for $Y(u_J)=10^3-10^2$. But if $Y(u_J)$ is indeed near the upper limit $10^3$ as seems the case for $u_J$'s of several km s$^{-1}$ \citep{Holsapple1993,Poelchauetal2013}, that would again only allow a limited margin of variation above solar values to avoid erosional destruction.

\subsection{Recoagulation downstream?}
\label{Recoagulation}

Assuming the protochondrule destruction condition is met, one could nevertheless expect that further downstream, when collision velocities allow sticking rather than disruption (below $\sim 10$ m s$^{-1}$), the dispersed droplets grow again. One may then wonder whether millimeter-size chondrules may be produced that way. 

  In his theoretical review on chondrule size, \citet{Jacquet2014size} found that this would require high particle densities. Indeed the density required to obtain a given size increase $\Delta a$ during a timescale $t_m$ (during which colliding droplets merge rather than bounce or freeze as compound chondrules) is \citep{Jacquet2014size}: 
\begin{eqnarray}
\label{Delta a}
\rho_d &=& 
\frac{\rho_0\Delta a}{\Delta v t_m} \nonumber\\
&=& 3\times 10^{-6}\:\mathrm{kg~m^{-3}}\left(\frac{\Delta a}{0.3\:\mathrm{mm}}\right)\left(\frac{10\:\rm m~s^{-1}}{\Delta v}\right)\left(\frac{10\:\rm h}{t_m}\right).\nonumber\\
\end{eqnarray}
Here $\Delta v$ is a measure of the (random) \footnote{The contribution of the \textit{systematic} velocity difference would be negligible. In fact, a calculation similar to that of Section \ref{Erosion}, if one puts $Y(v)=-1$, would replace the argument of the exponential with $-3\epsilon_1\Delta v/(4v_{T,2})\ll 1$ hence no appreciable change.} collision velocity. This is a very high density, comparable to the relatively high \textit{gas} densities adopted by \citet{MorrisDesch2010,Boleyetal2013}. 

  Note however that this would not necessarily prevent the shock models from accounting for the observed proportion of compound chondrules \citep{Ciesla2006,Morrisetal2012} if destruction was actually not extensive at the shock. The chondrule density $\rho_c$ corresponding to a compound fraction $x_{\rm cpd}$ is \citep{Jacquetetal2013}:
\begin{eqnarray}
\rho_c &=& \frac{2\rho_0a x_{\rm cpd}}{3\Delta v t_{\rm cpd}}\\
&=& 7\times 10^{-8}\:\mathrm{kg~m^{-3}}\left(\frac{x_{\rm cpd}}{4\: \%}\right)
{ ({\rho_0 a}/{\rm kg~m^{-2}}) \over (\Delta v/{\rm 10~m~s^{-1}})(t_{\rm cpd}/10~{\rm hr}) }\nonumber
\end{eqnarray}  
with $t_{\rm cpd}$ the time during which compound chondrule formation took place and where we have normalized $x_{\rm cpd}$ to the fraction found in ordinary chondrites \citep{GoodingKeil1981,Wassonetal1995}. 
If we adopt the higher end of conceivable sticking velocities \textit{and} compound formation times as in the above normalizations, this would be comparable to the solid densities in relatively dense disk models, with the help of the compression due to the shock itself \citep{Boleyetal2013}, although all these parameters are uncertain at the order-of-magnitude level.

  In principle, the emergence of the chondrules from the shocked zone may lead to further fragmentation. Indeed a pressure gradient over a lengthscale $\ell_{\rm cool}$ 
would incur a gas-grain drift (in the terminal velocity approximation of \citet{YoudinGoodman2005})
\begin{eqnarray}\label{eq:vpg}
v_{p-g} &=& \tau \frac{||\nabla P||}{\rho_g+\rho_p}\\
&\approx & c_g\frac{\rho_0a}{\left(\rho_g+\rho_p\right)\ell_{\rm cool}}\nonumber\\
&=& 0.2\:\mathrm{km~s^{-1}} \left(\frac{T}{2000\: \rm K}\right)^{1/2}\left(\frac{\rho_0a}{1\:\rm kg~m^{-2}}\right)\nonumber\\
&&\times \left(\frac{10^{-5}\:\rm kg~m^{-3}}{\rho_g+\rho_p}\right) \left(\frac{10^3\:\rm km}{\ell_{\rm cool}}\right).\nonumber
\end{eqnarray}
Here $c_g$ is the isothermal sound speed.  Equation (\ref{eq:vpg}) implies collisions at speeds well above the fragmentation limit as long as 
these collisions are not too frequent.  The number of collisions works out to  
\begin{equation}
\label{Ncoll emergence}
n_p4\pi a_p^2v_{p-g}\frac{\ell_{\rm cool}}{u_{g,2}}=3\sqrt{\frac{\pi}{8}}\theta_{\rm drag}\frac{\epsilon_p}{1+\epsilon_p}\frac{c_g}{u_{g,2}},
\end{equation}
which is $0.6\epsilon_p\ll 1$ for our (gas-dominated) shock parameters, hence little further fragmentation is to be expected during escape. 

Our general conclusion is that gas-dominated shocks, with heating regions extending further than a stopping length, 
would be generally detrimental to the integrity of chondrules. In the two sections to follow, we relax some of 
our assumptions to explore alternative regimes, namely solid-dominated shocks (Section \ref{Dusty shock}) and then 
small-scale bow shocks (Section \ref{Bow}).

\section{Structure of the Post-shock Flow with \\ High Particle Density}
\label{Dusty shock}

  Previous chondrule shock models have not considered large relative solid abundances. However, exploring such a regime is well motivated both by chondrule studies (as described in the Introduction) 
{\it and} by our developing understanding of protoplanetary disks.  These 
disks typically have a layered structure:  an outer magnetically active layer extends to a column $\sim 
10^2-10^3$ kg m$^{-2}$ below the disk surface \citep{Gammie1996}.    Particles that are initially suspended within a fully turbulent disk
(e.g. during an initial massive and gravitationally active phase) will mostly settle to the mid-plane once the
magnetorotational instability becomes the primary source of flow irregularities.\footnote{Gravity waves are excited in the
lower, laminar disk by the turbulent stresses above, but these waves do not interact effectively 
with small particles whose stopping time $\tau$ is much shorter than the orbital period.}  

Mass transfer in the disk may have the effect of raising the density of settled particles with respect to the gas, and solid abundances approaching the Roche density become possible once the vertically averaged abundance
ratio increases by a factor $\sim 10$ over solar values \citep[e.g.][]{YoudinShu2002}.  Beyond this point, planetesimal formation
becomes possible, and the particle layer is stirred by gravitationally driven modes. A planetesimal, or a large-scale shock, interacting with such a settled particle layer therefore encounters relatively well-defined conditions
at a distance of 2.5 AU from the Sun:  a total density approaching $\rho \sim 10^{-5}$ kg m$^{-3}$ and a solid/gas 
density ratio $\epsilon \sim \sqrt{2\pi}\rho c_g/\left(\Omega\Sigma_g\right) 
\sim  300(10^3~{\rm kg~m^{-2}}/\Sigma_g)$ with $\Omega$ the Keplerian angular velocity.  

In this section, we examine solid-dominated shocks taking into account the exchange of momentum and heat between particles and gas, the effects of fragmenting
collisions, and the emission and absorption of radiation by the particles.  The equations describing these
interactions, and the approximations made, are described in Appendix \ref{Dusty shock app}.  It is recalled that all plots in the paper, unless otherwise noted, have been produced using these more complete equations, with $10^{-3}<\epsilon_1<10^2$.

It is worth emphasizing that the processes described here do not depend
on a flow speed exceeding the {\it gas} sound speed.  That is because the
effective sound speed of waves with a period larger than the Epstein stopping
time (\ref{tstop}) of the embedded particles is
\begin{equation}
c_s = \left({\gamma P\over \rho_g + \rho_p}\right)^{1/2} = 
{c_{g,ad}\over(1+\epsilon)^{1/2}},
\end{equation}
where $c_{g,ad} = (\gamma P/\rho_g)^{1/2}$ is the usual adiabatic sound speed
in the gas.  A shock-like disturbance can form in a flow
with a speed exceeding $c_s$.  When $c_{g,ad} > u_1 > c_s$, the details
of the flow structure, and the characteristic fragmentation lengths,
will differ from the case $u_1 > c_{g,ad}$ in which a gas shock does form.
For example,  at large $\epsilon_1$, the 
effective thickness of the shock increases from the mean free path 
of gas particles to the mean free path $\sim \rho_0a/\rho_p$ for 
collisions between solid particles.  Nonetheless,
we find in practice that chondrule-melting temperatures generally require 
shock speeds larger than the gas sound speed at $T \sim 200$ K.  We
therefore maintain our focus on the flow behind a gas shock.

\subsection{Flow and particle behavior as a function of $\epsilon_1$}

\begin{figure}
\plotone{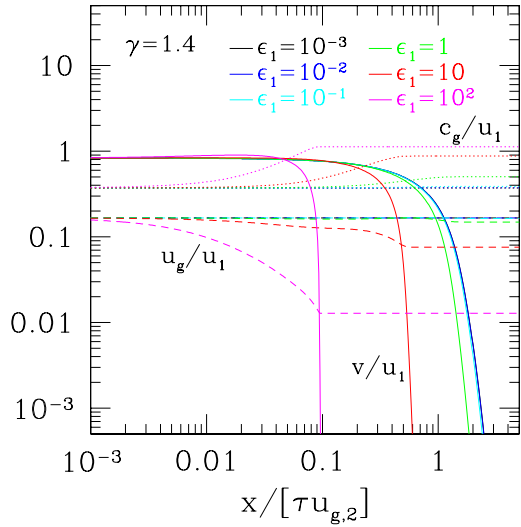}
\caption{Post-shock flow in a composite fluid of ideal gas (adiabatic index $\gamma = 1.4$) and solid particles with uniform stopping time $\tau$ upstream of the shock.  Solid lines:  particle velocity with respect to the gas; dashed lines:  gas velocity with respect to the shock; dotted lines:  gas sound speed.
Colors label the mean density $\epsilon_1$ of the solids relative to the gas, measured upstream of the shock.  Particle drag drives the rise in gas velocity and temperature post-shock, followed by a strong slow down at large $\epsilon_1$ as heat is absorbed from the gas.  The curves
for $\epsilon_1  = 10^{-3}$ and $\epsilon_1 = 10^{-2}$ do not differ measurably.  Radiation from the flow is ignored in this calculation.}
\vskip .2in
\label{fig:profile}
\end{figure}

\begin{figure}
\plotone{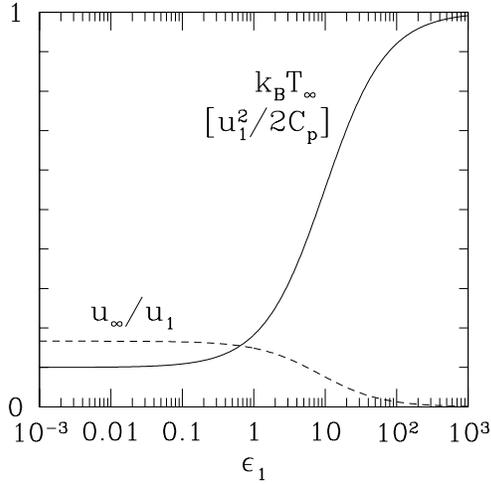}
\caption{Asymptotic temperature of solids in units of $u_1^2/2C_p$, for $\gamma = 1.4$, $C_p = 1.3$ kJ kg$^{-1}$K$^{-1}$,
$\mu_g = 2.3 m_u$ and vanishing ice density in the pre-shock flow.  
Dashed line shows the corresponding slow-down of the post-shock flow at large $\epsilon_1$.}
\label{fig:finaltemp}
\end{figure}

The flow structure for a single species of particle is shown in Figure \ref{fig:profile}, for several values of $\epsilon_1$, in the rest frame of the shock.  
Here collisions are not possible.  
The horizontal axis in the figure is normalized to the post-shock stopping length.

The mild superheating that the solids experience during their deceleration (Section \ref{Thermal evolution}) disappears for $\epsilon_1 \gtrsim 1$. One also observes a larger net compression of the flow for larger values of $\epsilon_1$, as a larger fraction of
the heat is stored in the solids, which contribute negligible pressure to the flow.   The asymptotic temperature
is also larger as a result of the lower specific heat of the solids.
Far from the shock, when gas and solids have equilibrated again both thermally and dynamically, the temperature
and flow speed relax to the values shown in Figure \ref{fig:finaltemp}.  The asymptotic expressions
\begin{equation}
\label{Tuinfty}
T_{\infty} \approx \frac{u_1^2}{2C_p}; \quad\quad u_\infty \approx \frac{k_Bu_1}{2\epsilon_1C_p\mu_g}
\end{equation}
apply when the solids dominate the specific heat ($\epsilon_1\gg C_g/2C_p \approx 10$). 
Temperatures in the chondrule-forming range can then be achieved for shock speeds as low as $\sim 2$ km s$^{-1}$. By comparison, for gas-dominated shocks, even a high gas density of $\rho_{g,1}=10^{-6}$ kg m$^{-3}$ would require a $u_1$ of about 8 km s$^{-1}$ (see Equation (\ref{Tpeak})), amounting to some 40 \% of the orbital speed at 2.5 AU. Moreover, the reduction in $u_{g,\infty}$ would allow longer cooling times for small-scale shocks such as those produced by planetary embryos.

\begin{figure}
\plotone{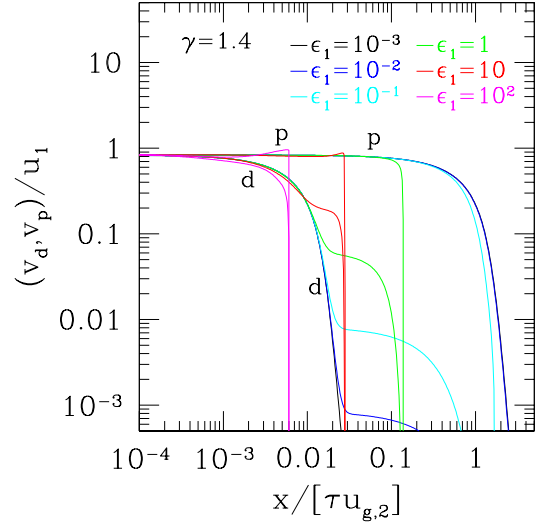}
\caption{Post-shock velocity profile of two species of particles, with size ratio $a_d/a_p = 0.01$.  Collisions between
the particles result in mass ejection from large particles according to Equation (\ref{Y}), assuming $v_f = 0.04\,u_1$ and $\alpha = 1.5$. 
All fragments have same size $a_d$, which does not evolve, and their mutual collisions are neglected.   Radiation from the flow is handled using the treatment described in Appendix \ref{Dusty shock app}, assuming $\kappa_{\rm rad} = 10$.}
\vskip .2in
\label{fig:profile2}
\end{figure}

\begin{figure}
\plotone{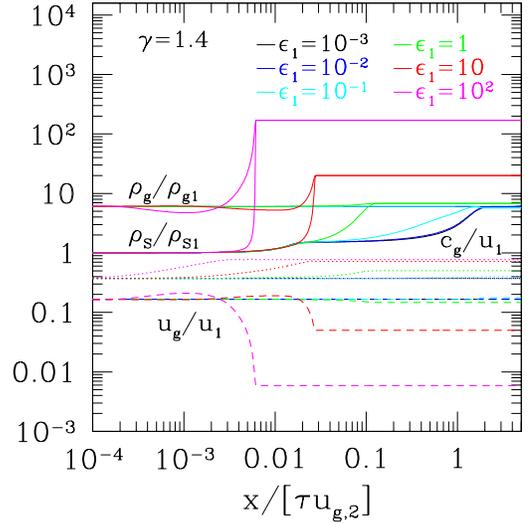}
\caption{Post-shock profiles of density (both gas and net solid density, denoted by $\rho_S$),
gas velocity and gas sound speed corresponding to Figure \ref{fig:profile2}.}
\vskip .2in
\label{fig:profile3}
\end{figure}

\begin{figure}
\plotone{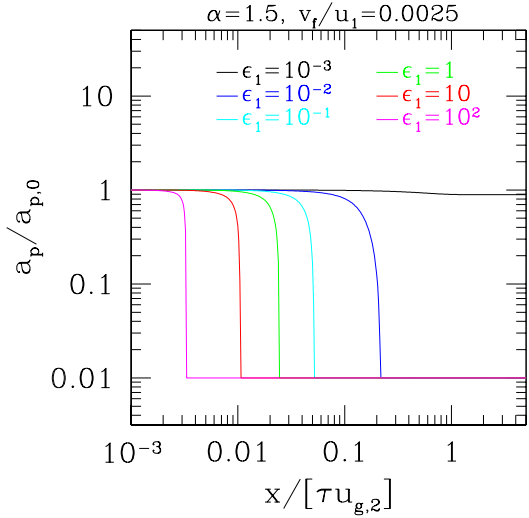}
\caption{Profile of the size of large particles (solid lines), corresponding to Figure \ref{fig:profile2} but now with
$v_f$ reduced to $0.0025 u_1$.
}
\vskip .2in
\label{size}
\end{figure}

\begin{figure}
\plotone{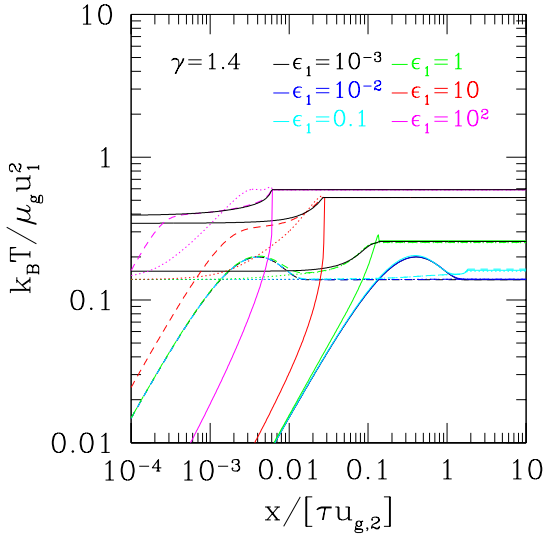}
\caption{Temperature of solids (dashed lines: small particles; solid lines: large particles), gas (dotted lines), and radiation field
(defined as $(E_{\rm rad}/a)^{1/4}$, thin black lines, shown for $\epsilon_1 = 1$-$10^2$).  Corresponding flow profiles shown in
Figures \ref{fig:profile2} and \ref{fig:profile3}.}
\vskip .2in
\label{fig:temp}
\end{figure}

The differential deceleration of large and small particles is shown in Figure \ref{fig:profile2} (see also Figures \ref{fig:profile3}, \ref{size} and \ref{fig:temp}).  Here we include the effect
of fragmenting collisions using the prescription for erosion given in Equation (\ref{Y}), and consider an initial
particle size ratio $a_s/a_b = 10^{-2}$.  The smaller `dust' particles are assumed to be fixed in size.  The larger
particles experience faster deceleration as they are ablated, leading to a sharp destruction layer.  

Radiative energy losses are negligible on the stopping length of the particles in the case of small particle loadings;
but they become significant for large $\epsilon_1$. 
The optical depth in dust also rises sufficiently (Appendix \ref{Dusty shock app}
) to begin to trap heat in the flow,
as the incident particles are breaking down into smaller grains.  
This means that only a modest fraction of the incident kinetic energy flux is lost to radiation back across the shock
(see Figure \ref{fig:radeff} in Appendix \ref{Dusty shock app}).  The
strong compression of the flow occuring at larger $\epsilon_1$ is somewhat enhanced by radiative energy losses.



\subsection{Chondrule survival}

Here we assess the survival of chondrules in solid-dominated shocks as compared to gas-dominated shocks. As may be seen in Figure \ref{size}, solid-dominated shocks also lead to wholesale destruction of protochondrules by sandblasting, as may have been intuitively expected. This is despite the reduced time ($\propto \left(\rho_p+\rho_g\right)^{-1}$) for dynamical equilibration (because of feedback on the gas and recoil during collisions) which does not reverse the effect of an increasing collision rate ($\propto \rho_p$).  

 In contrast with gas-dominated shocks, there is now a good prospect for recoagulation. 
If the total pre-shock density is $\sim 10^{-5}$ kg~m$^{-3}$, as limited by gravitational stability, and we take
into account the compression behind the shock as enhanced by heat exchange between solids and gas (see above), 
downstream densities up to $\rho_p \sim 10^{-4}$ kg m$^{-3}$ may be considered. 
This would provide rapid enough collisions to reproduce chondrule size as may be judged from Equation (\ref{Delta a}).

The collision rate would also be enhanced during escape (see Equation (\ref{Ncoll emergence})), but now
the collision speed between particles $v_{p-p}$ would be limited by the high frequency of collisions.
By balancing the acceleration driven by gas pressure gradients against the effect of mutual collisions, we have 
\begin{equation}
4\pi a^2 n_p v_{p-p}^2\sim {v_{p-g}\over \tau}\sim \frac{\rho_g}{\rho_g+\rho_p}\frac{c_g^2}{\ell_{\rm cool}}.
\end{equation}
This gives
\begin{eqnarray}
v_{p-p} &\approx & c_g\epsilon_1^{-1/2}\left({\rho_0 a\over 3\rho_p \ell_{\rm cool}}\right)^{1/2}\nonumber\\
&\approx & \left(\frac{\rho_0ak_B}{3t_{\rm cool}\mu_g\epsilon_1\rho_{p,1}}\right)^{1/2}\left(\frac{T_{\infty}}{2C_p}\right)^{1/4}\nonumber\\
&=& 10\:\mathrm{m~s^{-1}}\left(\frac{10^2}{\epsilon_1}\right)^{1/2}\left(\frac{10^{-5}\:\rm kg~m^{-3}}{\rho_{p,1}}\right)^{1/2}\nonumber\\
&&\times \left(\frac{\rho_0a}{1\:\rm kg~m^{-2}}\right)^{1/2}
\left(\frac{10\:\rm h}{t_{\rm cool}}\right)^{1/2}\left(\frac{T_{\infty}}{2000\:\rm K}\right)^{1/4},\nonumber\\
\end{eqnarray}
where we have set $l_{\rm cool}=u_\infty t_{\rm cool}$ and made use of the asymptotic Equations (\ref{Tuinfty}). Coherent solids (as opposed to aggregates) would have fragmentation velocities of order 10-100 m s$^{-1}$ \citep{Vedderetal1974} and the destructive effect of collisions between solid chondrules would also be reduced by the repeated accretion and ejection of dusty envelopes.



  A problem remains with having chondrules form by recoagulation:  their composition would be expected to be homogenized, 
inconsistent with observations \citep{HezelPalme2007,Jacquet2014size}.  Some variation could be reproduced if the final assembly of 
the reformed chondrules involved particles of intermediate size, which recorded stochastically different ranges of temperature and 
density in the flow, and therefore experienced variable exchange with the gas.  A significant challenge for such a scenario
is however provided by the measured dispersion in the abundance of refractory elements like rare earths 
\citep{Jacquet2014size}.

\section{Flash heating in small \\ planetesimal bow shocks}
\label{Bow}

  The preceding calculations are predicated on the assumption that the shocked region is larger than the particle 
stopping length.  One could envision small-scale shocks, specifically bow shocks around planetesimals, that are smaller than $u_J\tau_2$ and 
yet thermally affect solids.  Since the heating timescale is only one order of magnitude below the stopping time, this would essentially 
require the whole thermal episode to last $\sim t_{\rm th}$.

  In this case, solid destruction would be diminished from the estimates of the previous sections.   Collisions between 
protochondrules would be somewhat rarer, in spite of being also possible after exiting the shocked region because of the relative 
velocities developed by then (see appendix \ref{Wake}).  Sandblasting would also be reduced in the shocked region (see last paragraph in 
Section \ref{Erosion}), and also further downstream since the ejection yield there $\sim Y(u_1t_{\rm th}/\tau)$ would be reduced 
(see Equation (\ref{dm/dv})).

 Such a thermal episode would be quite short.  If we eliminate $\rho_{g,1}$ using the expression for $T_{\rm peak}$ in 
 Equation (\ref{Tpeak}), we have: 
\begin{eqnarray}
\label{t_th calculated}
t_{\rm th}&=& \frac{2}{3}\frac{\gamma-1}{\left(\gamma+1\right)^2}\frac{\mu_gC_p\rho_0a u_1^2}{k_B\sigma_{\rm SB}e_mT_{\rm peak}^4}\\
&=& 1.5\:\mathrm{s}\left(\frac{u_1}{8\:\rm km~s^{-1}}\right)^2\left(\frac{\rho_0a}{1\:\rm kg~m^{-2}}\right)\left(\frac{2000\:\rm K}{T_{\rm peak}}\right)^4\left(\frac{0.8}{e_m}\right).\nonumber
\end{eqnarray}
This is much shorter than cooling timescales (hours-days) generally assigned to chondrules from furnace experiments, trace element and 
crystallographic systematics \citep[][and references therein]{Hewinsetal2005,Jacquetetal2012CC}. \citet{Rubin2000,WassonRubin2003} 
nevertheless argue for such short cooling times to avoid loss of Na, which would be complete within $\sim 10$ s under canonical 
nebular conditions \citep[e.g.][]{FedkinGrossman2013}. While a single thermal pulse sufficiently intense to melt the precusor solids 
would produce textures very different from known chondrules \citep{Hewinsetal2005}, \citet{WassonRubin2003} argue for a \textit{succession} 
of less intense flash heatings (at least tens) to achieve gradual growth of crystals.  We note that, in that case, the loss of Na would be 
the integrated sum of that of each of those events and would actually still overpredict the observations. 
While chondrules may not then have been produced by such flash heating events, we may still consider the formation of agglomeratic olivine objects \citep{WeisbergPrinz1996,Ruzickaetal2012} or igneous rims around 
chondrules \citep{KrotWasson1995} by such a process. 

 How important would such planetesimal-induced flash heating be in the early solar system? Since the extent of the bow shock is comparable to the planetesimal radius (about 3 times in the calculations of \citet{Cieslaetal2004}), the planetesimal size required would be of order: 
 \begin{eqnarray}
\label{lengthscale}
u_1t_{\rm th}
&=& \frac{2}{3}\frac{\gamma-1}{\left(\gamma+1\right)^2}\frac{\mu_gC_p\rho_0a u_1^3}{k_B\sigma_{\rm SB}e_mT_{\rm peak}^4}\\
&=& 12\:\mathrm{km}\left(\frac{u_1}{8\:\rm km~s^{-1}}\right)^3\left(\frac{\rho_0a}{1\:\rm kg~m^{-2}}\right)\left(\frac{2000\:\rm K}{T_{\rm peak}}\right)^4\left(\frac{0.8}{e_m}\right)\nonumber
\end{eqnarray} 
Interestingly, the cross sectional area of the \textit{present-day} asteroid belt appears dominated by asteroids around 6 km in diameter \citep{Ivezicetal2001}, about the appropriate size range, although the primordial planetesimal size distribution may have been different \citep[e.g.][]{Morbidellietal2009}.

  The solid mass $\Delta M_{\rm proc}$ that is processed in one orbit by a given planetesimal of radius $R_{\rm pl}$ and mass $M_{\rm pl}$ is given by
\begin{eqnarray}
\label{bow efficiency}
\frac{\Delta M_{\rm proc}}{M_{\rm pl}}&=&\frac{3}{4}\theta_{\rm orb}\left(1+z\right)\frac{\Sigma_p}{\rho_0R_{\rm pl}}\\
&=& 10^{-6}\left(\frac{\theta_{\rm orb}}{2}\right)\left(1+z\right)\left(\frac{\Sigma_p}{10\:\rm kg~m^{-2}}\right)\left(\frac{5\:\rm km}{R_{\rm pl}}\right).
\nonumber
\end{eqnarray}
Here $z$ is the relative increase in the processing cross section over the geometrical one, $\Sigma_p$ is the surface density of solids in the local disk annulus, and $\theta_{\rm orb}$ a order unity number\footnote{For an inclined circular orbit, $\theta_{\rm orb}=2/\mathrm{cos}(i/2)$ with $i$ the inclination ($\gtrsim \mathrm{arcsin}(H/R)$); for an eccentric in-plane orbit, $\theta_{\rm orb} \approx 4 \langle u_1\rangle /v_{T,1}$
.}. 

Over a few Ma, depending on the fraction of adequately excited few km-sized planetesimals in the planetesimal belt, some significant 
processing could be achieved. Independently of damping timescales \citep{Morrisetal2012,HoodWeidenschilling2012}, the ``flash-heating'' 
activity of a given excited planetesimal would be limited by erosion by millimeter-size solids.  Indeed, any such solid with impact 
parameter $< R_{\rm pl}$ would impact the planetesimal as gas drag would not have time to significantly alter their 
trajectories \citep{Morrisetal2012} and they would do so at almost full velocity $u_1$, with high ejection yields 
$10^2-10^3$ \citep{Poelchauetal2013}.  The resulting lifetime works out to be $\ll 10^5$ a (as can be seen by 
multiplying the right-hand side of Equation (\ref{bow efficiency}) by $Y(u_1)/(1+z)$). 

\section{Summary and conclusions}
\label{Discussion}

We have investigated, analytically and numerically, the fate of solids in shocks during the deceleration phase.

 We found that, in the gas-dominated regime, the deceleration of solids could be reasonably well described with a constant stopping time $\tau$, lending itself to analytic treatment. Melting of the protochondrules would occur at about one tenth this stopping time, but evaporation of supermicron-size dust grains would not be completed within the deceleration timescale, especially for supersolar solid/gas ratios. We argued that collisions at velocities greater than 10-100 m s$^{-1}$ will lead to fragmentation and/or splashing.

 We found that solid/gas ratios $\gtrsim 10^{-1}$ would lead to high-speed disrupting collisions for protochondrules with moderate size 
dispersions (factor of $\sim 2$) comparable to those observed for chondrules in individual chondrites 
\citep[e.g.][]{NelsonRubin2002,Teitleretal2010}.  Sandblasting of protochondrules by dust (rapidly coupled to the gas), whether inherited 
from the pre-shock region or produced through fragmentation in the post-shock region, may be even more efficient at destroying protochondrules.
Indeed the ejecta produced by such collisions enhance the dust concentration, generating a positive feedback.  
Melting of chondrule surfaces may alter the effects of sandblasting; but even so large particles have generally been generally broken down before melting due to rapid collisions. 

As a result, 
the threshold solid/gas ratio for wholesale destruction scales inversely with the ejection yield at the postshock gas-solid velocity, and might be 
as small as the solar value.  
In the standard gas-dominated shock models, there is little prospect that once protochondrules are destroyed, recoagulation further downstream can reconstitute mm-sized particles, although collisions would be frequent enough to reproduce the statistics of compound chondrules for reasonable parameters, if chondrules did escape wholesale destruction. 

We then considered solid-dominated shocks.  Such a setting, which may more easily satisfy the empirical constraints from chondrules,
may be astrophysically conceivable in the framework of layered accretion where solids remain settled in a dead zone and thus gradually 
concentrate relative to the gas.  Solid-dominated shocks would require smaller velocities than gas-dominated shocks to achieve
chondrule-melting temperatures, and would more easily produce long cooling times (for small-scale shocks) and retention of volatiles.  While destruction of chondrules past the shock would be extensive, subsequent recoagulation would be possible because of the higher densities -- although assembling chondrules that way may conflict with their observed compositional diversity.
 
  We also considered small-scale bow shocks due to few km-sized planetesimals where deceleration is incomplete. While destruction of solids would be limited compared to the above, the hereby flash-heated products (timescales of a few seconds) would likely not resemble chondrules but these may possibly account for slightly melted features in chondrites such as agglomeratic olivine objects or some igneous rims.

  So it would appear that existing nebular shock wave models for chondrule formation are inconsistent with significant enhancements of the solid/gas above solar values. 
 Nonetheless, such enrichments are suggested by several chondrule observations such as the frequency of compound 
chondrules \citep{GoodingKeil1981,Cieslaetal2004}, FeO content \citep{Grossmanetal2012,Schraderetal2013}, 
and Na retention \citep{Alexanderetal2008}.  Such observations would need to be explained in another way 
for the shock model to remain viable.  We have already seen that there is some room to adjust the parameters to 
account for the compound chondrule fraction.  Such may also be marginally the case for the FeO constraint of type I 
(FeO-poor) chondrules, and type II (FeO-rich) chondrules could be envisioned to have formed by a mechanism 
qualitatively different from that of their type I counterparts. 
One option, especially with regard to the Na retention issue, is to invoke a nonsolar 
chemistry for the gas component e.g. due to the outgassing of a planetary atmosphere, as proposed by \citet{Morrisetal2012} 
for the bow shock model.  (This conclusion has yet to be verified quantitatively as the atmosphere in question would 
be far below the shock front.)  At any rate, such a solution cannot be applied to large-scale shocks such as those 
expected from gravitational instabilities---in that it would seem that bow shocks are favored \citep[see also][]{StammlerDullemond2014}. 
But unless such alternatives to a global enrichment of the chondrule-forming environment can be found, and ejection yields during erosion 
are shown to be $\lesssim 10^2$ at several km s$^{-1}$, then destruction of chondrules remains a serious issue for shock wave scenarios.

\bibliographystyle{aa}
\bibliography{bibliography}

\begin{thebibliography}{80}
\expandafter\ifx\csname natexlab\endcsname\relax\def\natexlab#1{#1}\fi

\bibitem[{{Akaki} \& {Nakamura}(2005)}]{AkakiNakamura2005}
{Akaki}, T. \& {Nakamura}, T. 2005, GCA, 69, 2907

\bibitem[{{Alexander} {et~al.}(2008){Alexander}, {Grossman}, {Ebel}, \&
  {Ciesla}}]{Alexanderetal2008}
{Alexander}, C. M.~O., {Grossman}, J.~N., {Ebel}, D.~S., \& {Ciesla}, F.~J.
  2008, Science, 320, 1617

\bibitem[{{Birnstiel} {et~al.}(2010){Birnstiel}, {Dullemond}, \&
  {Brauer}}]{Birnstieletal2010}
{Birnstiel}, T., {Dullemond}, C.~P., \& {Brauer}, F. 2010, A\&A, 513, A79+

\bibitem[{{Boley} {et~al.}(2013){Boley}, {Morris}, \& {Desch}}]{Boleyetal2013}
{Boley}, A.~C., {Morris}, M.~A., \& {Desch}, S.~J. 2013, ApJ, 776, 101

\bibitem[{{Boss}(1996)}]{Boss1996}
{Boss}, A.~P. 1996, in Chondrules and the Protoplanetary Disk, ed.
  {R.~H.~Hewins, R.~H.~Jones, \& E.~R.~D.~Scott}, 257--263

\bibitem[{{Boss} \& {Durisen}(2005{\natexlab{a}})}]{BossDurisen2005}
{Boss}, A.~P. \& {Durisen}, R.~H. 2005{\natexlab{a}}, ApJL, 621, L137

\bibitem[{{Boss} \& {Durisen}(2005{\natexlab{b}})}]{BossDurisen2005sources}
{Boss}, A.~P. \& {Durisen}, R.~H. 2005{\natexlab{b}}, in Astronomical Society
  of the Pacific Conference Series, Vol. 341, Chondrites and the Protoplanetary
  Disk, ed. A.~N. {Krot}, E.~R.~D. {Scott}, \& B.~{Reipurth} (Astronomical
  Society of the Pacific), 821

\bibitem[{{Brauer} {et~al.}(2008){Brauer}, {Dullemond}, \&
  {Henning}}]{Braueretal2008}
{Brauer}, F., {Dullemond}, C.~P., \& {Henning}, T. 2008, A\&A, 480, 859

\bibitem[{{Brearley} \& {Jones}(1998)}]{BrearleyJones1998}
{Brearley}, A. \& {Jones}, A. 1998, in Planetary Materials, ed. J.~J. {Papike}
  (Mineralogical Society of America), 3--1--3--398

\bibitem[{{Bridges} {et~al.}(2012){Bridges}, {Changela}, {Nayakshin},
  {Starkey}, \& {Franchi}}]{Bridgesetal2012}
{Bridges}, J.~C., {Changela}, H.~G., {Nayakshin}, S., {Starkey}, N.~A., \&
  {Franchi}, I.~A. 2012, EPSL, 341, 186

\bibitem[{{Ciesla}(2005)}]{Ciesla2005}
{Ciesla}, F.~J. 2005, in Astronomical Society of the Pacific Conference Series,
  Vol. 341, Chondrites and the Protoplanetary Disk, ed. {A.~N.~Krot,
  E.~R.~D.~Scott, \& B.~Reipurth}, 811--820

\bibitem[{{Ciesla}(2006)}]{Ciesla2006}
{Ciesla}, F.~J. 2006, M\&PS, 41, 1347

\bibitem[{{Ciesla} {et~al.}(2004{\natexlab{a}}){Ciesla}, {Hood}, \&
  {Weidenschilling}}]{Cieslaetal2004bow}
{Ciesla}, F.~J., {Hood}, L.~L., \& {Weidenschilling}, S.~J. 2004{\natexlab{a}},
  M\&PS, 39, 1809

\bibitem[{{Ciesla} {et~al.}(2004{\natexlab{b}}){Ciesla}, {Lauretta}, \&
  {Hood}}]{Cieslaetal2004}
{Ciesla}, F.~J., {Lauretta}, D.~S., \& {Hood}, L.~L. 2004{\natexlab{b}}, M\&PS,
  39, 531

\bibitem[{{Connolly} \& {Desch}(2004)}]{ConnollyDesch2004}
{Connolly}, Jr., H.~C. \& {Desch}, S.~J. 2004, Chemie der Erde/Geochemistry,
  64, 95

\bibitem[{{Cuzzi} {et~al.}(2001){Cuzzi}, {Hogan}, {Paque}, \&
  {Dobrovolskis}}]{Cuzzietal2001}
{Cuzzi}, J.~N., {Hogan}, R.~C., {Paque}, J.~M., \& {Dobrovolskis}, A.~R. 2001,
  ApJ, 546, 496

\bibitem[{{Desch}(2007)}]{Desch2007}
{Desch}, S.~J. 2007, ApJ, 671, 878

\bibitem[{{Desch} {et~al.}(2005){Desch}, {Ciesla}, {Hood}, \&
  {Nakamoto}}]{Deschetal2005}
{Desch}, S.~J., {Ciesla}, F.~J., {Hood}, L.~L., \& {Nakamoto}, T. 2005, in
  Astronomical Society of the Pacific Conference Series, Vol. 341, Chondrites
  and the Protoplanetary Disk, ed. {A.~N.~Krot, E.~R.~D.~Scott, \&
  B.~Reipurth}, 849--871

\bibitem[{{Desch} \& {Connolly}(2002)}]{DeschConnolly2002}
{Desch}, S.~J. \& {Connolly}, Jr., H.~C. 2002, M\&PS, 37, 183

\bibitem[{{Desch} {et~al.}(2012){Desch}, {Morris}, {Connolly}, \&
  {Boss}}]{Deschetal2012}
{Desch}, S.~J., {Morris}, M.~A., {Connolly}, H.~C., \& {Boss}, A.~P. 2012,
  M\&PS, 47, 1139

\bibitem[{{Fedkin} \& {Grossman}(2013)}]{FedkinGrossman2013}
{Fedkin}, A.~V. \& {Grossman}, L. 2013, GCA, 112, 226

\bibitem[{{Gammie}(1996)}]{Gammie1996}
{Gammie}, C.~F. 1996, ApJ, 457, 355

\bibitem[{{Gombosi} {et~al.}(1986){Gombosi}, {Nagy}, \&
  {Cravens}}]{Gombosietal1986}
{Gombosi}, T.~I., {Nagy}, A.~F., \& {Cravens}, T.~E. 1986, Rev. Geophys., 24,
  667

\bibitem[{{Gooding} \& {Keil}(1981{\natexlab{a}})}]{GoodingKeil1981viscosity}
{Gooding}, J.~L. \& {Keil}, K. 1981{\natexlab{a}}, in Lunar and Planetary Inst.
  Technical Report, Vol.~12, Lunar and Planetary Institute Science Conference
  Abstracts, 353--355

\bibitem[{{Gooding} \& {Keil}(1981{\natexlab{b}})}]{GoodingKeil1981}
{Gooding}, J.~L. \& {Keil}, K. 1981{\natexlab{b}}, Meteoritics, 16, 17

\bibitem[{{Grossman} {et~al.}(2012){Grossman}, {Fedkin}, \&
  {Simon}}]{Grossmanetal2012}
{Grossman}, L., {Fedkin}, A.~V., \& {Simon}, S.~B. 2012, M\&PS, 47, 2160

\bibitem[{{Guignard}(2011)}]{Guignard2011}
{Guignard}, J. 2011, PhD thesis, Universit\'{e} Toulouse III

\bibitem[{{G{\"u}ttler} {et~al.}(2010){G{\"u}ttler}, {Blum}, {Zsom}, {Ormel},
  \& {Dullemond}}]{Guettleretal2010}
{G{\"u}ttler}, C., {Blum}, J., {Zsom}, A., {Ormel}, C.~W., \& {Dullemond},
  C.~P. 2010, A\&A, 513, A56

\bibitem[{{Hayashi}(1981)}]{Hayashi1981}
{Hayashi}, C. 1981, Progress of Theoretical Physics Supplement, 70, 35

\bibitem[{{Hewins} {et~al.}(2005){Hewins}, {Connolly}, \&
  {Libourel}}]{Hewinsetal2005}
{Hewins}, R.~H., {Connolly}, Lofgren, G.~E. J. H.~C., \& {Libourel}, G. 2005,
  in Astronomical Society of the Pacific Conference Series, Vol. 341,
  Chondrites and the Protoplanetary Disk, ed. {A.~N.~Krot, E.~R.~D.~Scott, \&
  B.~Reipurth}, 286--316

\bibitem[{{Hewins} {et~al.}(2012){Hewins}, {Zanda}, \&
  {Bendersky}}]{Hewinsetal2012}
{Hewins}, R.~H., {Zanda}, B., \& {Bendersky}, C. 2012, GCA, 78, 1

\bibitem[{{Hezel} \& {Palme}(2007)}]{HezelPalme2007}
{Hezel}, D.~C. \& {Palme}, H. 2007, GCA, 71, 4092

\bibitem[{{Holsapple}(1993)}]{Holsapple1993}
{Holsapple}, K.~A. 1993, Annu. Rev. Earth Planet. Sci., 21, 333

\bibitem[{{Hood} \& {Horanyi}(1991)}]{HoodHoranyi1991}
{Hood}, L.~L. \& {Horanyi}, M. 1991, Icarus, 93, 259

\bibitem[{{Hood} \& {Weidenschilling}(2012)}]{HoodWeidenschilling2012}
{Hood}, L.~L. \& {Weidenschilling}, S.~J. 2012, M\&PS, 47, 1715

\bibitem[{{Housen} \& {Holsapple}(2011)}]{HousenHolsapple2011}
{Housen}, K.~R. \& {Holsapple}, K.~A. 2011, Icarus, 211, 856

\bibitem[{{Housen} {et~al.}(1983){Housen}, {Schmidt}, \&
  {Holsapple}}]{Housenetal1983}
{Housen}, K.~R., {Schmidt}, R.~M., \& {Holsapple}, K.~A. 1983, J. Geophys.
  Res., 88, 2485

\bibitem[{{Howard}(1802)}]{Howard1802}
{Howard}, E. 1802, Philosophical Transaction, 92, 168

\bibitem[{{Iida} {et~al.}(2001){Iida}, {Nakamoto}, {Susa}, \&
  {Nakagawa}}]{Iidaetal2001}
{Iida}, A., {Nakamoto}, T., {Susa}, H., \& {Nakagawa}, Y. 2001, Icarus, 153,
  430

\bibitem[{{Ivezi{\'c}} {et~al.}(2001){Ivezi{\'c}}, {Tabachnik}, {Rafikov},
  {Lupton}, {Quinn}, {Hammergren}, {Eyer}, {Chu}, {Armstrong}, {Fan},
  {Finlator}, {Geballe}, {Gunn}, {Hennessy}, {Knapp}, {Leggett}, {Munn},
  {Pier}, {Rockosi}, {Schneider}, {Strauss}, {Yanny}, {Brinkmann}, {Csabai},
  {Hindsley}, {Kent}, {Lamb}, {Margon}, {McKay}, {Smith}, {Waddel}, {York}, \&
  {SDSS Collaboration}}]{Ivezicetal2001}
{Ivezi{\'c}}, {\v Z}., {Tabachnik}, S., {Rafikov}, R., {et~al.} 2001,
  Astronomical Journal, 122, 2749

\bibitem[{{Jacquet}(2014)}]{Jacquet2014size}
{Jacquet}, E. 2014, Icarus, 232, 176

\bibitem[{{Jacquet} {et~al.}(2012){Jacquet}, {Alard}, \&
  {Gounelle}}]{Jacquetetal2012CC}
{Jacquet}, E., {Alard}, O., \& {Gounelle}, M. 2012, M\&PS, 47, 1695

\bibitem[{{Jacquet} {et~al.}(2013){Jacquet}, {Paulhiac-Pison}, {Alard}, \&
  {Kearsley}}]{Jacquetetal2013}
{Jacquet}, E., {Paulhiac-Pison}, M., {Alard}, O., \& {Kearsley}, A. 2013,
  M\&PS, 48, 1981

\bibitem[{{Kadono} {et~al.}(2008){Kadono}, {Arakawa}, \&
  {Kouchi}}]{Kadonoetal2008}
{Kadono}, T., {Arakawa}, M., \& {Kouchi}, A. 2008, Icarus, 197, 621

\bibitem[{Kring(1991)}]{Kring1991}
Kring, D.~A. 1991, EPSL, 105, 65

\bibitem[{{Krot} {et~al.}(2009){Krot}, {Amelin}, {Bland}, {Ciesla}, {Connelly},
  {Davis}, {Huss}, {Hutcheon}, {Makide}, {Nagashima}, {Nyquist}, {Russell},
  {Scott}, {Thrane}, {Yurimoto}, \& {Yin}}]{Krotetal2009}
{Krot}, A.~N., {Amelin}, Y., {Bland}, P., {et~al.} 2009, GCA, 73, 4963

\bibitem[{{Krot} \& {Wasson}(1995)}]{KrotWasson1995}
{Krot}, A.~N. \& {Wasson}, J.~T. 1995, GCA, 59, 4951

\bibitem[{{Lodders}(2003)}]{Lodders2003}
{Lodders}, K. 2003, ApJ, 591, 1220

\bibitem[{{Miura} \& {Nakamoto}(2005)}]{MiuraNakamoto2005}
{Miura}, H. \& {Nakamoto}, T. 2005, Icarus, 175, 289

\bibitem[{{Morbidelli} {et~al.}(2009){Morbidelli}, {Bottke}, {Nesvorn{\'y}}, \&
  {Levison}}]{Morbidellietal2009}
{Morbidelli}, A., {Bottke}, W.~F., {Nesvorn{\'y}}, D., \& {Levison}, H.~F.
  2009, Icarus, 204, 558

\bibitem[{{Morris} \& {Desch}(2014)}]{MorrisDesch2014}
{Morris}, M. \& {Desch}, S. 2014, in 45th Lunar and Planetary Science
  Conference, 2577

\bibitem[{{Morris} {et~al.}(2012){Morris}, {Boley}, {Desch}, \&
  {Athanassiadou}}]{Morrisetal2012}
{Morris}, M.~A., {Boley}, A.~C., {Desch}, S.~J., \& {Athanassiadou}, T. 2012,
  ApJ, 752, 27

\bibitem[{{Morris} \& {Desch}(2010)}]{MorrisDesch2010}
{Morris}, M.~A. \& {Desch}, S.~J. 2010, ApJ, 722, 1474

\bibitem[{{Nagasawa} {et~al.}(2014){Nagasawa}, {Tanaka}, {Tanaka}, {Nakamoto},
  {Miura}, \& {Yamamoto}}]{Nagasawaetal2014}
{Nagasawa}, M., {Tanaka}, K.~K., {Tanaka}, H., {et~al.} 2014, ArXiv e-prints

\bibitem[{{Nakamoto} \& {Miura}(2004)}]{NakamotoMiura2004}
{Nakamoto}, T. \& {Miura}, H. 2004, in Lunar and Planetary Institute Science
  Conference Abstracts, Vol.~35, Lunar and Planetary Institute Science
  Conference Abstracts, ed. S.~{Mackwell} \& E.~{Stansbery}, 1847

\bibitem[{{Nelson} \& {Rubin}(2002)}]{NelsonRubin2002}
{Nelson}, V.~E. \& {Rubin}, A.~E. 2002, M\&PS, 37, 1361

\bibitem[{{Planchette} {et~al.}(2012){Planchette}, {Lorenceau}, \&
  {Brenn}}]{Planchetteetal2012}
{Planchette}, C., {Lorenceau}, E., \& {Brenn}, G. 2012, Journal of Fluid
  Mechanics, 702, 5

\bibitem[{{Poelchau} {et~al.}(2013){Poelchau}, {Kenkmann}, {Thoma}, {Hoerth},
  {Dufresne}, \& {Sch{\'n}Fer}}]{Poelchauetal2013}
{Poelchau}, M.~H., {Kenkmann}, T., {Thoma}, K., {et~al.} 2013, M\&PS, 48, 8

\bibitem[{{Qian} \& {Law}(1997)}]{QianLaw1997}
{Qian}, J. \& {Law}, C.~K. 1997, Journal of Fluid Mechanics, 331, 59

\bibitem[{{Rein}(1993)}]{Rein1993}
{Rein}, M. 1993, Fluid Dynamics Research, 12, 61

\bibitem[{{Roscoe}(1952)}]{Roscoe1952}
{Roscoe}, R. 1952, British Journal of Applied Physics, 3, 267

\bibitem[{{Rubin}(2000)}]{Rubin2000}
{Rubin}, A.~E. 2000, Earth Science Reviews, 50, 3

\bibitem[{{Ruzicka} {et~al.}(2012){Ruzicka}, {Floss}, \&
  {Hutson}}]{Ruzickaetal2012}
{Ruzicka}, A., {Floss}, C., \& {Hutson}, M. 2012, GCA, 76, 103

\bibitem[{{Schrader} {et~al.}(2013){Schrader}, {Connolly}, {Lauretta},
  {Nagashima}, {Huss}, {Davidson}, \& {Domanik}}]{Schraderetal2013}
{Schrader}, D.~L., {Connolly}, H.~C., {Lauretta}, D.~S., {et~al.} 2013, GCA,
  101, 302

\bibitem[{{Stammler} \& {Dullemond}(2014)}]{StammlerDullemond2014}
{Stammler}, S.~M. \& {Dullemond}, C.~P. 2014, ArXiv e-prints

\bibitem[{{Susa} \& {Nakamoto}(2002)}]{SusaNakamoto2002}
{Susa}, H. \& {Nakamoto}, T. 2002, ApJL, 564, L57

\bibitem[{{Tachibana} \& {Huss}(2005)}]{TachibanaHuss2005}
{Tachibana}, S. \& {Huss}, G.~R. 2005, GCA, 69, 3075

\bibitem[{{Teitler} {et~al.}(2010){Teitler}, {Paque}, {Cuzzi}, \&
  {Hogan}}]{Teitleretal2010}
{Teitler}, S.~A., {Paque}, J.~M., {Cuzzi}, J.~N., \& {Hogan}, R.~C. 2010,
  M\&PS, 45, 1124

\bibitem[{{Tsuchiyama} {et~al.}(1999){Tsuchiyama}, {Tachibana}, \&
  {Takahashi}}]{Tsuchiyamaetal1999}
{Tsuchiyama}, A., {Tachibana}, S., \& {Takahashi}, T. 1999, GCA, 63, 2451

\bibitem[{{Uesugi} {et~al.}(2005){Uesugi}, {Akaki}, {Sekiya}, {Nakamura},
  {Tsuchiyama}, {Nakano}, \& {Uesugi}}]{Uesugietal2005}
{Uesugi}, M., {Akaki}, T., {Sekiya}, M., {et~al.} 2005, in Astronomical Society
  of the Pacific Conference Series, Vol. 341, Chondrites and the Protoplanetary
  Disk, ed. A.~N. {Krot}, E.~R.~D. {Scott}, \& B.~{Reipurth}, 893

\bibitem[{{Vedder} {et~al.}(1974){Vedder}, {Gault}, \&
  {Larimer}}]{Vedderetal1974}
{Vedder}, J.~F., {Gault}, D.~E., \& {Larimer}, J.~W. 1974, Science, 185, 378

\bibitem[{{Walzel}(1980)}]{Walzel1980}
{Walzel}, P. 1980, Chem. Ing. Tech., 52, 338

\bibitem[{{Wasson} {et~al.}(1995){Wasson}, {Krot}, {Min}, \&
  {Rubin}}]{Wassonetal1995}
{Wasson}, J.~T., {Krot}, A.~N., {Min}, S.~L., \& {Rubin}, A.~E. 1995, GCA, 59,
  1847

\bibitem[{{Wasson} \& {Rubin}(2003)}]{WassonRubin2003}
{Wasson}, J.~T. \& {Rubin}, A.~E. 2003, GCA, 67, 2239

\bibitem[{{Weisberg} \& {Prinz}(1996)}]{WeisbergPrinz1996}
{Weisberg}, M.~K. \& {Prinz}, M. 1996, in Chondrules and the Protoplanetary
  Disk, ed. R.~H. {Hewins}, R.~H. {Jones}, \& E.~R.~D. {Scott}, 119--127

\bibitem[{{Wood}(1963)}]{Wood1963chondrules}
{Wood}, J.~A. 1963, Icarus, 2, 152

\bibitem[{{Yarin}(2006)}]{Yarin2006}
{Yarin}, A.~L. 2006, Annual Reviews of Fluid Mechanics, 38, 159

\bibitem[{{Youdin} \& {Goodman}(2005)}]{YoudinGoodman2005}
{Youdin}, A.~N. \& {Goodman}, J. 2005, ApJ, 620, 459

\bibitem[{{Youdin} \& {Shu}(2002)}]{YoudinShu2002}
{Youdin}, A.~N. \& {Shu}, F.~H. 2002, ApJ, 580, 494

\bibitem[{{Zsom} {et~al.}(2010){Zsom}, {Ormel}, {G{\"u}ttler}, {Blum}, \&
  {Dullemond}}]{Zsometal2010}
{Zsom}, A., {Ormel}, C.~W., {G{\"u}ttler}, C., {Blum}, J., \& {Dullemond},
  C.~P. 2010, A\&A, 513, A57

\end{thebibliography}

\begin{appendix}

\section{Calculating Post-shock Flow with Finite $\epsilon_1$}\label{Dusty shock app}

Here we outline the method used to calculate the flow profiles shown in Section \ref{Dusty shock}, as well as the mean collision
speeds in Figures \ref{collision speed}, \ref{collision speed2}.

In the frame of the shock,  the speed of a particle is written as $u = v + u_g$,  where $v$ is the differential
speed used in Sections \ref{Single} and \ref{Collisions}.  
The conservation of momentum and energy can be written -- in the limit $M_1 \gg 1$ --  as
\begin{equation}\label{eq:momint}
P_g + \rho_g u_g \left[u_g + \epsilon_1 u\right]  = (1+\epsilon_1) \rho_1 u_1^2,
\end{equation}
\begin{equation}\label{eq:eint}
{1\over 2}u_g^2 + {\gamma\over\gamma-1}{P_g\over\rho_g} + \epsilon_1\left({1\over 2}u^2 + C_pT_p\right) = 
(1+\epsilon_1){1\over 2}u_1^2
\end{equation}
(see also \citealt{MorrisDesch2010}).  These equations are easily generalized to multiple particle species by
weighting by the mass fraction $f_i$ in particle species $i$ ($\sum_i f_i = 1$):
\begin{equation}\label{eq:uav}
u \rightarrow \overline{u} = \sum_i f_i u_i; \quad u^2 \rightarrow \overline{u^2} = \sum_i f_i u_i^2,
\end{equation}
and analogously for solid temperature.
The cumulative effect of radiation from the post-shock flow is incorporated by changing the right-hand side of
Equation (\ref{eq:eint}) to 
\begin{equation}
(1+\epsilon_1){1\over 2}u_1^2 \rightarrow f_{\rm rad} \cdot (1+\epsilon_1){1\over 2}u_1^2 ,
\end{equation}
where $f_{\rm rad} \leq 1$.

The velocity of a particle of species $i$ evolves according to
\begin{eqnarray}\label{eq:fdrag}
u_i{du_i\over dx} = -{(u_i-u_g)\over \tau(a_i)} 
- \sum_{i\neq j}{u_i-u_j\over m_j + m_i} |u_i-u_j| \pi (a_i + a_j)^2 \rho_j,
\end{eqnarray}
where $m_i = (4\pi/3)\rho_0 a_i^3$ is the mass of particle species $i$, $a_i$ its radius, and $\rho_i = n_i m_i$ the 
corresponding mean density.   Collisions develop
between particles of different sizes because they are subject to differing gas drag.  The second term in equation
(\ref{eq:fdrag}) is normalized in the approximation where collisions are elastic.

When implementing sand-blasting, we make the approximation of a single species of small particle with a 
single velocity at each $x$.  There is an additional exchange of momentum from large particles to dust that
is associated with the growth in $\rho_d$,
\begin{equation}
\epsilon_d {du_d\over dx} \rightarrow \epsilon_d{du_d\over dx} - \sum_i {d\epsilon_i\over dx}(u_i-u_d).
\end{equation}

The temperature(s) of the particles follow equation  (\ref{eq:Tp})  with parameters $T_{\rm rec}$ and $\theta_{\rm th}$ 
evaluated following \cite{Gombosietal1986}.   The gas velocity can be reconstructed by substituting the particle
averages (\ref{eq:uav}) into the integrals (\ref{eq:momint}) and (\ref{eq:eint}), suitably modified for radiative energy loss:
\begin{equation}
{u_g\over u_1} =  {\gamma[1 - X_1(\overline{u}/u_1)] - D^{1/2}\over (\gamma+1)(1-X_1)},
\end{equation}
with $X_1=\epsilon_1/(1+\epsilon_1)$ the fraction of the mass flux carried by particles, and where
\begin{eqnarray}
D \equiv \gamma^2[1 - X_1(\overline{u}/u_1)]^2 
-  (\gamma^2-1)(1-X_1)\left[f_{\rm rad} - {X_1\over u_1^2}\left(\overline{u^2} + 2C_p\overline{T_s}\right)\right].
\end{eqnarray}
The gas density is $\rho_g = \rho_{g,1} (u_1/u_g)$, and 
\begin{equation}
{c_g\over u_1} = {(u_g/u_1)^{1/2}\over (1-X_1)^{1/2}}\left[1 - (1-X_1)(u_g/u_1) - X_1(\overline{u_s}/u_1)\right]^{1/2}.
\end{equation}

The radiation field itself is handled by using the Eddington closure approximation.  A steady flux of radiation interacting
with $N$ species of particles of variable size satisfies the equation
\begin{equation}
{dF\over dx} = \sum_{j=1}^N\alpha_j[4\sigma_{\rm SB}T_{s,j}^4 - cE_{\rm rad}],	
\end{equation}
and the closure condition
\begin{equation}
{1\over 3}{dE_{\rm rad}\over dx} = -F\sum_{j=1}^N \alpha_j.
\end{equation}
We set the emissivities to the geometric values; hence the absorption coefficients are given by
\begin{equation}
\label{alpha_j}
\alpha_j \equiv \pi a_j^2 n_j = {3\epsilon_j\rho_g\over 4\rho_0 a_j}.
\end{equation}

The amplitude of radiative losses is conveniently parameterized by
\begin{eqnarray}
k_{\rm rad} = {\sigma_{\rm SB}\left({\mu_g u_1^2/k_B}\right)^4 \over {1\over2}(1+\epsilon_1)\rho_{g,1}u_1^3}
     = 2.1\left(u_1\over 2~{\rm km~s^{-1}}\right)^5\left[\frac{10^{-5}\: \rm kg~m^{-3}}{(1+\epsilon_1)\rho_{g,1}}\right],
\end{eqnarray}
which is taken to be $k_{\rm rad} = 10$ in our calculations of flows with radiation.

In the calculations presented here, we choose simple boundary conditions on the radiation field:  the outgoing flux
at the shock has magnitude $F = cE_{\rm rad}/\sqrt{3}$, and $F = 0$ at the point in the flow where large particles have broken down into
small grains by sandblasting.  This is, of course, only an approximate procedure for small values of $\epsilon_1$,
but in that case the flow structure is insensitive to radiation transport on the scale where the solids are breaking down.	

\begin{figure}
\includegraphics[width=0.5\textwidth]{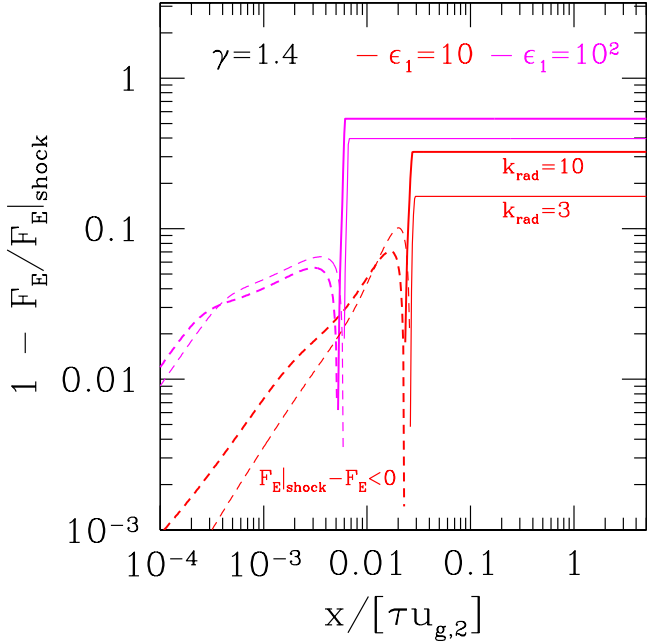}
\caption{Combined kinetic and thermal energy flux of the solids and gas, relative to the incoming kinetic energy flux,
in the flow solutions with large solid abundance ($\epsilon_1 = 10,10^2$) shown in Figures \ref{fig:profile2}, \ref{fig:profile3}.
The difference well downstream of the shock represents the outgoing radiation energy flux.}
\vskip .2in
\label{fig:radeff}
\end{figure}

\begin{figure}
\includegraphics[width=0.5\textwidth]{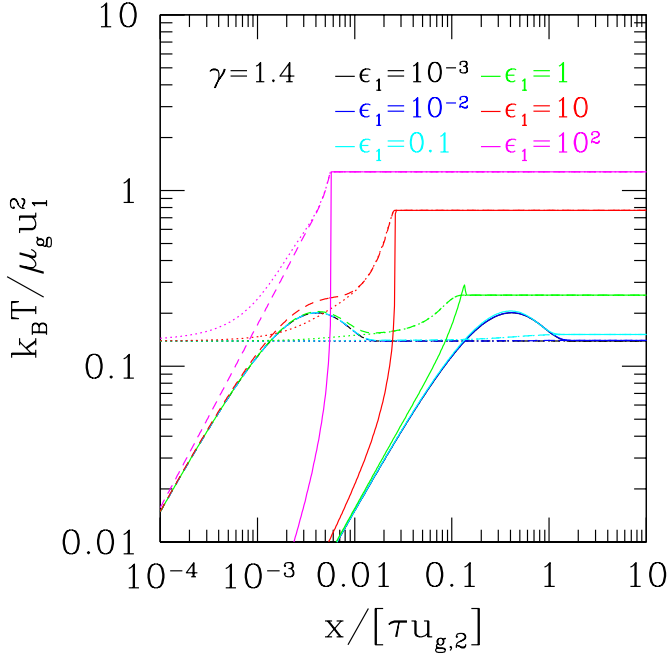}
\caption{Post-shock temperature profile in a flow identical to that of Figure \ref{fig:temp}, but with radiative emission
and absorption by the particles suppressed.} 
\vskip .2in
\label{fig:temp2}
\end{figure}

\subsection{Effect of radiation from the solids on the flow profile}

Radiation has only a small effect on the calculated flow profiles.   At small $\epsilon_1$, that is because we only follow
the flow over the relatively small stopping length of the particles.   The reason is different at large $\epsilon_1$:  there
sandblasting of the larger particles creates a significant optical depth in small grains, which then provide significant
trapping of radiation. Indeed, from Equation (\ref{dmdt}), ignoring the variation of the relative velocity $v\approx u_J$, a sandblasting length (in the rest frame of the shock) may be estimated as
\begin{equation}
\ell_{\rm sand}=\frac{4\rho_0a_pu_J}{3\rho_dY(u_J)u_1},
\end{equation} 
from which an optical depth of the destruction layer of
\begin{eqnarray}\label{eq:taug}
\tau_{\rm geom}= \int_0^{\ell_{\rm sand}} n_d \pi a_d^2 \mathrm{d}x
\approx  \frac{1}{2Y(u_J)}\frac{a_p}{a_d}\frac{u_J}{u_1}
=\left(\frac{10^2}{Y(u_J)}\right)\left(\frac{a_p}{0.3\:\rm mm}\right)\left(\frac{1\:\rm \mu m}{a_d}\right)
\end{eqnarray}
may be obtained.  This result is fairly insensitive to grain fragmentation given the $\sim a_d/\lambda$ dependence of the absorption 
coefficient for submicron grains, hence our use of 
the geometric value in the numerical calculations (Equation (\ref{alpha_j})). 

  The outgoing radiation flux at the shock is only a moderate fraction of the incoming kinetic
energy flux of solids and gas (Figure \ref{fig:radeff}).  Radiation is effectively trapped in flows with $\epsilon_1 = 10, 10^2$.  

The profiles of velocity and particle size shown in Sections \ref{Collisions} and \ref{Dusty shock} are essentially unchanged
if radiative cooling is removed from the flow.  A larger difference is apparent in the gas and solid temperature profiles:
the result shown in Figure \ref{fig:temp2} may be compared with Figure \ref{fig:temp}.

\section{Collisions in the wake of a small bow shock}
\label{Wake}

In this appendix, we consider the case where the shocked region ``2'' has a finite length $L\ll u_1\tau_s$ (in the rest frame of the shock). 
We want to estimate the limiting value for the initial separation $l_1$ of a big and a small particle for mutual collision either 
(i) in the shocked region; or (ii) after exiting the shocked region (as relative velocities will have already developed).  We crudely approximate the adiabatic expansion of the heated gas into the ambient medium by considering 
that after exiting, the (heretofore uniform) gas properties (velocities, temperature, densities) 
suddenly return to their pre-shock value.  We call this final region ``3''.

\subsection{Single-particle dynamics in region 2}

  For a given particle, the time spent in the shocked region ``2'' is given by:
\begin{equation}
L=u_{g,2}t_2+u_J\tau_2\left(1-e^{-t_2/\tau_2}\right)\approx u_1t_2-\frac{u_Jt_2^2}{2\tau_2}+o\left(\left(\frac{t_2}{\tau_2}\right)^2\right)
\end{equation}
which may be solved approximately as\footnote{Although only the first-order (in $L$) term will matter in the final result, keeping the second-order one will allow the inquisitive reader to check that the corrections will be indeed negligible in the second part of the appendix.}:
\begin{equation}
t_2\approx\frac{L}{u_1}\left(1+\frac{Lu_J}{2u_1^2\tau_2}\right).
\end{equation}
Hence the (negative) velocity relative to the gas upon entry in region ``3'' is:
\begin{eqnarray}
v_{\rm 3, in}= u_J\left(e^{-t_2/\tau_2}-1\right)=\frac{u_{g,2}t_2-L}{\tau_2}
\approx-\frac{u_JL}{u_1\tau_2}\left(1-\frac{u_{g,2}L}{2u_1^2\tau_2}\right)
\end{eqnarray}

\subsection{Collision of a pair in region 3}

  We place ourselves in the rest frame of the gas in region ``3''. For a given pair of big and small particle, the distance between the two points of entry in region ``3'' is:
\begin{equation}
l_3=u_1\left(\frac{l_1}{u_1}+t_{b,2}-t_{s,2}\right)\approx l_1+\frac{L^2u_J}{2u_1^3}\left(\frac{1}{\tau_{b,2}}-\frac{1}{\tau_{s,2}}\right).
\end{equation}
Hence, if we take the new origin of times and abscissas to correspond to the entry of the small particle in region ``3'', the difference of abscissas of the small and the big particle are, for $t>\mathrm{max}(0,l_3/u_1)\equiv t_{\rm start,3}$:
\begin{eqnarray}
x_s-x_b=v_{\rm s,3, in}\tau_{s,3}\left(1-e^{-t/\tau_{s,3}}\right)
-\left(-l_3+v_{\rm b,3, in}\tau_{b,3}\left(1-e^{-(t-l_3/u_1)/\tau_{b,3}}\right)\right)
\end{eqnarray}
As a function of $t\in \mathbb{R}$, this reaches a minimum for
\begin{equation}
t_{\rm min}=\frac{1}{\tau_{s,3}^{-1}-\tau_{b,3}^{-1}}\left(\mathrm{ln}\left(\frac{v_{\rm s,3,in}}{v_{\rm b,3,in}}\right)-\frac{l_3}{u_1\tau_{b,3}}\right),
\end{equation}
which minimum evaluates to
\begin{eqnarray}
\label{minimum}
x_s(t_{\rm min})-x_b(t_{\rm min})=l_3+u_{g,2}\frac{\tau_{b,3}}{\tau_{b,2}}\left(t_{s,2}-t_{b,2}\right)
+e^{\frac{l_3}{u_1(\tau_{b,3}-\tau_{s,3})}}v_{\rm b,3,in}\tau_{b,3}\left(\frac{v_{\rm b,3,in}}{v_{\rm s, 3, in}}\right)^{\frac{1}{\tau_{b,3}/\tau_{s,3}-1}}\left(1-\frac{\tau_{s,3}}{\tau_{b,3}}\right),
\end{eqnarray}
before the abscissa difference starts to increase, then converging toward
\begin{equation}
\label{limit}
l_3+u_{g,2}\frac{\tau_{b,3}}{\tau_{b,2}}\left(t_{s,2}-t_{b,2}\right)=l_1+u_1\left(\left(\frac{T_{g,2}}{T_{g,3}}\right)^{1/2}-1\right)\left(t_{s,2}-t_{b,2}\right)
\end{equation}
as $t\longrightarrow +\infty$. Physically, this change of direction of variation (which is not present in the infinite shocked region) 
indicates that the big particle first experiences a headwind of small particles and then a tailwind (because the small particles have coupled more rapidly to the gas moving at speed $u_1$ in the rest frame of the shock). 
So, in principle, collisions are conceivable for $l_1>0$ \textit{and} $l_1<0$.

  Let us consider the case $l_1>0$ first. 
\vskip .1in
\noindent 1. If a collision took place in the shocked region ``2'', $l_3<0$ and the initial abscissa difference $x_s-x_b$ is negative and so is (\textit{a fortiori}) its minimum. 
\vskip .1in
\noindent 2. If no collision took place in region ``2'' ($l_3>0$) but is to take place in region ``3'', the minimum must be reached for $t_{\rm min}>l_3/u_1\Leftrightarrow l_3<u_1\tau_{s,3}\mathrm{ln}(v_{\rm s,3,in}/v_{\rm b,3,in})$ and be negative.
\vskip .1in
The two subcases can thus be synthesized in the necessary condition \textit{that $l_3<u_1\tau_{s,3}\mathrm{ln}(v_{\rm s,3,in}/v_{\rm b,3,in})$ and the minimum (\ref{minimum}) is negative}. 

  Given that $l_1>0\Rightarrow\lim\limits_{t \to +\infty}(x_s-x_b)>0$ , it can be seen that the italicized conditions, implying that the (negative) minimum is reached after $t_{\rm start,3}$ (as even if $l_3<0$, $t_{\rm min}>0=t_{\rm start,3}$ here), actually imply (except when the minimum is exactly zero) \textit{two} distinct formal collision times (since even if the initial $x_s-x_b$ is already negative (first subcase) and allow only one collision in region 3, we must have had another one in the shocked region). We will come back to this formal plurality later, but are here content to conclude that the italicized conditions above are sufficient (as well as necessary). In the limit $L\ll u_1\tau_s$, it can be shown that they amount to the single inequality  
\begin{equation}
\label{lcrit bis}
l_1<l_{\rm crit, bis}\equiv\left(\frac{a_s}{a_b}\right)^{\frac{1}{a_b/a_s-1}}\left(1-\frac{a_s}{a_b}\right)\frac{u_J}{u_1}\frac{\tau_{b,3}}{\tau_{b,2}}L.
\end{equation}

  The case $l_1<0$ (implying $l_3<0$ and hence a negative initial $x_s-x_b$) is simpler. A (one-time) overtaking of the big particle by the small particle is indeed possible if the limit (\ref{limit}) of the abscissa difference for $t\longrightarrow +\infty$ is positive, which yields the condition
\begin{eqnarray}
 l_1>l_{\rm min,bis}\equiv-u_1\left(\left(\frac{T_{g,2}}{T_{g,3}}\right)^{1/2}-1\right)\left(t_{s,2}-t_{b,2}\right)
&\approx& -\frac{u_JL^2}{2u_1^2\tau_{s,2}}\left(\left(\frac{T_{g,2}}{T_{g,3}}\right)^{1/2}-1\right)\left(1-\frac{a_s}{a_b}\right)\nonumber\\
&=&-\left(\frac{a_b}{a_s}\right)^{\frac{1}{a_b/a_s-1}}\left(\frac{\rho_{g,2}}{\rho_{g,3}}-\frac{\tau_{b,2}}{\tau_{b,3}}\right)\left(\frac{L}{2u_1\tau_{s,2}}\right)l_{\rm crit,bis}.\nonumber
\end{eqnarray}

In the case of dynamically cold dust, the nominal mean collision number (that is ignoring scattering), to which the collision probability 
is related, would be $\left(l_{\rm crit,bis}-l_{\rm min,bis}\right)/l_{\rm coll,1}$ since the two collisions times for $0\leq l_1<l_{\rm crit,bis}$ would 
only count as one (the small particle, if existent at $l_1$ would have been scattered by the first collision).  However, in a realistic dust, the section 
dust tailwind nominally partly emptied (or found empty) by the big particle (in its ``headwind'' phase) will have been replenished, so that the real 
collision number would be $(2l_{\rm crit,bis}-l_{\rm min,bis})/l_{\rm coll,1}\approx 2l_{\rm crit, bis}/l_{\rm coll,1}$.\footnote{For very small 
particles (but keeping $L\ll u_1\tau_b$), we would have the condition $-(u_JL/u_{g,2})((T_{g,2}/T_{g,3})^{1/2}-1)<l_1<u_JL/u_{g,2}$, with the positive $l_1$ being 
swept twice, so ``back-sweeping'' upstream would be here important.} 
 Its ratio to the collision number $l_{\rm crit}/l_{\rm coll, 1}$ in the standard scenario with full deceleration is thus
\begin{equation}\label{eq:last}
2\frac{l_{\rm crit,bis}}{l_{\rm crit}}\approx 2\left(\frac{a_s}{a_b}\right)^{\frac{1}{a_b/a_s-1}}\left(\frac{T_{g,2}}{T_{g,3}}\right)^{1/2}\left(\frac{t_{b,2}}{\tau_{b,2}}\right).
\end{equation} 
If the temperature contrast does not exceed (say) one order of magnitude, then the new condition is indeed more stringent than before (the second factor in Equation (\ref{eq:last}) varies between $e^{-1}$ and 1). 
This gain may be limited if $t_{b,2}\sim t_{\rm th}=0.15\tau_{b,2}$, but it 
does push $\epsilon_{\rm s,crit}$ for commensurately sized bodies closer to unity. 

\end{appendix}

\end{document}